\documentclass[twocolumn,showpacs,nofootinbib,preprintnumbers,amsmath,amssymb,showkeys]{revtex4}

\def\Tr{{\rm Tr}}
\newcommand{\avg}[1]{\left< #1 \right>} 
\usepackage{enumitem}
\usepackage{epsfig}
\usepackage{dsfont}
\usepackage{color}
\usepackage{graphicx}
\usepackage{dcolumn}
\usepackage{bm}
\usepackage{amsmath}

\def\mathrlap{\mathpalette\mathrlapinternal}
\def\mathrlapinternal#1#2{%
           \rlap{$\mathsurround=0pt#1{#2}$}}

\def\slashii#1{\setbox0=\hbox{$#1$}             
   \dimen0=\wd0                                 
   \setbox1=\hbox{\sl/} \dimen1=\wd1            
   \ifdim\dimen0>\dimen1                        
      \rlap{\hbox to \dimen0{\hfil\sl/\hfil}}   
      #1                                        
   \else                                        
      \rlap{\hbox to \dimen1{\hfil$#1$\hfil}}   
      \hbox{\sl/}                               
   \fi}                                         %
\def\slashiii#1{\setbox0=\hbox{$#1$}#1\hskip-\wd0\hbox to\wd0{\hss\sl/\/\hss}}
\newcommand\pfrac[3]{\left ( \frac{#1}{#2} \right )^{#3} }
\newcommand{\mysinh}[3]{\left [  \bigg(\frac{#1}{#2}\bigg)^{\mathrlap{\!\!#3}} \,\, -\, \bigg(\frac{#2}{#1}\bigg)^{\!\!#3} \right]}
\newcommand{\mycosh}[3]{\left [ \bigg(\frac{#1}{#2}\bigg)^{\mathrlap{\!\!#3}} \, \,+ \,\bigg(\frac{#2}{#1}\bigg)^{\!\!#3} \right]}
\newcommand{\eq}[1]{\begin{align} #1 \end{align}}
\newcommand{\beq}{\begin{equation}}
\newcommand{\eeq}{\end{equation}}
\newcommand{\bea}{\begin{eqnarray}}
\newcommand{\eea}{\end{eqnarray}}
\newcommand{\nn}{\nonumber \\}

\newcommand\eqn[1]{(\ref{#1})}      
\newcommand\Eqn[1]{Eq.~(\ref{#1})}  
\newcommand\Fig[1]{Fig.~\ref{#1}}  

\begin{document}


\title{Nonperturbative renormalization group for scalar fields \\in de Sitter space: Beyond the local potential approximation}

\author{Maxime Guilleux}

\author{Julien Serreau}

\affiliation{APC, AstroParticule et Cosmologie, Universit\'e Paris Diderot, CNRS/IN2P3, CEA/Irfu, Observatoire de Paris, Sorbonne Paris Cit\'e \\
10, rue Alice Domon et L\'eonie Duquet, 75205 Paris Cedex 13, France
}

\date{\today}

\begin{abstract}

Nonperturbative renormalization group techniques have recently proven a powerful tool to tackle the nontrivial infrared dynamics of light scalar fields in de Sitter space. In the present article, we develop the formalism beyond the local potential approximation employed in earlier works. In particular, we consider the derivative expansion, a systematic expansion in powers of field derivatives, appropriate for long wavelength modes, that we generalize to the relevant case of a curved metric with Lorentzian signature. The method is illustrated with a detailed discussion of the so-called local potential approximation prime which, on top of the full effective potential, includes a running (but field-independent) field renormalization. We explicitly compute the associated anomalous dimension for O($N$) theories. We find that it can take large values along the flow, leading to sizable differences as compared to the local potential approximation. However, it does not prevent the phenomenon of gravitationally induced dimensional reduction pointed out in previous studies. We show that, as a consequence, the effective potential at the end of the flow is unchanged as compared to the local potential approximation, the main effect of the running anomalous dimension being merely to slow down the flow.

 \end{abstract}

\maketitle

\section{Introduction}

The inflationary paradigm provides a consistent picture of the early Universe with both observational and theoretical successes. It has also brought much attention to the topic of quantum fields in curved space-time and, more specifically, to the case of de Sitter space. Of particular interest is the case of light scalar fields (in units of the space-time curvature), whose quantum fluctuations undergo a dramatic amplification on superhorizon scales due to the accelerated expansion. This can be viewed as a tremendous particle production from the classical gravitational field~\cite{Mottola:1984ar}. If this is responsible for the observed power spectrum of primordial density fluctuations in inflationary cosmology \cite{Parentani:2004ta}, it also results in a nontrivial infrared dynamics \cite{Starobinsky:1994bd,Tsamis:2005hd,Weinberg:2005vy}. In particular, perturbative (loop) contributions are typically plagued by infrared and secular divergences which require resummation techniques or genuine nonperturbative approaches \cite{Starobinsky:1994bd,Mazzitelli:1988ib,Burgess:2009bs,Rajaraman:2010xd,Serreau:2011fu,Beneke:2012kn,Akhmedov:2011pj,Gautier:2013aoa,Boyanovsky:2012qs,Serreau:2013psa,Serreau:2013koa,Nacir:2013xca,Nacir:2016fzi}. Among those, the nonperturbative renormalization group (NPRG; see Refs.~\cite{Berges:2000ew,Delamotte:2007pf} for reviews), recently adapted to the case of scalar fields in de Sitter space \cite{Kaya:2013bga,Serreau:2013eoa,Guilleux:2015pma}, provides a promising tool to study the onset of gravitational effects as one progressively integrates modes from subhorizon to superhorizon scales. 

Nontrivial results have been obtained in this context for O($N$) scalar field theories using the simple local potential approximation (LPA)~\cite{Serreau:2013eoa,Guilleux:2015pma}, where one only retains the full functional flow of the effective potential but neglects that of other (e.g. derivative) terms~\cite{Morris:1994ki}. In particular, it has been shown that the large superhorizon fluctuations induce a dimensional reduction of the renormalization group (RG) flow, resulting in an effective zero-dimensional theory. The effective potential can be expressed in terms of a normal integral---as opposed to a functional one---which can be put in exact correspondence with the late time stationary probability distribution function of the stochastic approach \cite{Starobinsky:1994bd}. This dimensional reduction also nicely explains the phenomenon of radiative symmetry restoration discussed in Ref.~\cite{Ratra:1984yq} for the case $N=2$ and in Refs.~\cite{Mazzitelli:1988ib,Serreau:2011fu} for $N\to\infty$. The NPRG analysis shows that such gravitationally induced symmetry restoration occurs for any value of $N$ and in any space-time dimension, in agreement with the results of Ref.~\cite{Lazzari:2013boa} in the stochastic approach. The NPRG approach has recently been applied to the study of symmetry restoration in scalar quantum electrodynamics in de Sitter space, with similar results \cite{Gonzalez:2016jrn}.

It is, therefore, of interest to investigate possible corrections to the LPA, which is the simplest, yet nontrivial, approximation in the context of NPRG methods. A typical extension is the derivative expansion, where one includes the running of the kinetic term and of higher derivative terms in the effective action~\cite{Morris:1994ie}. This is suitable for the study of physical quantities primarily sensitive to long wavelength field configurations, such as critical exponents at a continuous phase transition in statistical physics \cite{Berges:2000ew,Delamotte:2007pf}. Other approximation schemes can be based on a functional expansion of the effective action in powers of the fields. This generates an infinite tower of coupled flow equations for vertex functions which has to be truncated in one way or another. In the present work, we undertake the study of NPRG methods in de Sitter space beyond the LPA, focusing on the derivative expansion, having in mind the dynamics of long wavelength fluctuations. Our aim is twofold: First, establishing a consistent formulation of the derivative expansion in de Sitter space; second, putting the formalism at work in the simplest {\it {\it Ansatz}} beyond the LPA.

To this aim, we shall investigate the so-called local potential approximation prime (LPA'), where one includes a running, but field-independent renormalization of the standard kinetic term. This allows for relatively simpler calculations as compared to the full (field-dependent) next-to-leading order in the derivative expansion. In the context of statistical physics, the LPA' is able to capture the main qualitative features of the phase structure of O($N$) scalar theories in flat Euclidean space. In particular, it correctly describes the existence of a nontrivial critical regime for the two-dimensional XY model ($N=2$), which corresponds to the Berezinsky-Kosterlitz-Thouless (BKT) transition \cite{Berezinsky:1970fr,Kosterlitz:1973xp,Kosterlitz:1974sm,Berges:2000ew,Delamotte:2007pf}. For the present purposes, this approximation is enough to illustrate the approach. We shall see that, due to the phenomenon of dimensional reduction described above, the inclusion of the running field renormalization factor alters the RG trajectories but not the end result of the flow as far as the effective potential is concerned. For instance, symmetry restoration happens later along the flow---i.e., deeper in the infrared---but the final effective potential exactly agrees with that of the LPA and, thus, with the stochastic approach.

The paper is organized as follows. In Sec.~\ref{sec:setup}, we present the general NPRG setup in de Sitter space-time. We discuss the issues of a proper formulation of the derivative expansion in general curved space-time and present a consistent approach in de Sitter space-time. We apply the formalism to the LPA' for O($N$) scalar theories in Secs.~\ref{sec:N1} and \ref{sec:multiple}. We present a detailed calculation of the running anomalous dimension and we discuss the infrared and ultraviolet limits. The resulting flow equations in the infrared regime and the consequences of the gravitational dimensional reduction are discussed in Sec.~\ref{sec:discussion} and we conclude in Sec.~\ref{sec:conclusion}. Some technical details are presented in Appendixes~\ref{appsec:correlators}, \ref{appsec:IRlimit}, and \ref{ap:flow} and we present a discussion of the LPA' flow equations in Minkowski space-time with particular emphasis on the Lorentz violating effects from the regulator in Appendix~\ref{appsec:Mink}.

\section{General setting}\label{sec:setup}

\subsection{NPRG in de Sitter space}

The basic setup of the NPRG approach in de Sitter space has been developed in Refs.~\cite{Kaya:2013bga,Serreau:2013eoa,Guilleux:2015pma}. We briefly review it here.
We consider the expanding Poincar\'e patch of a de Sitter space-time with Lorentzian signature in $D=d+1$ dimensions.  In terms of the  comoving spatial coordinates ${\bf X}$ the line element reads
\eq{
 ds^2 = -dt^2+e^{2Ht}d{\bf X}^2=a^2(\eta) \left(-d\eta^2 +d{\bf X}^2\right),
}
where the cosmological time $t\in\mathds{R}$ and the conformal time $\eta\in\mathds{R}^-$ are related by $a(\eta) = -1/(H\eta)=e^{Ht}$, with $H$ the expansion rate. Unless explicitly stated, we set $H=1$ in the following. We shall be interested in the case of scalar field theories and, to fix the ideas, we shall explicitly consider theories with O($N$) symmetry described by the classical action
\begin{equation}
S[\varphi] = -\int_x \left\{\frac{1}{2} \partial_\mu\varphi_a\partial^\mu\varphi_a+V(\varphi)  \right\},
\end{equation}
where  $\int_x = \int d^Dx\sqrt{-g(x)}=\int d\eta \,a^D(\eta)\int d^dX $ is the invariant integration measure, with $g(x)$ the determinant of the metric tensor. The potential $V(\varphi)$ is a function of the O($N$) invariant $\varphi_a\varphi_a$, and a summation over repeated space-time or O($N$) indices $a=1,\ldots,N$ is understood. Note that the potential $V(\varphi)$ includes possible couplings to the (constant) space-time curvature.

The NPRG approach consists in deforming the classical action with a space-time-dependent mass term, $S\to S_\kappa = S + \Delta S_\kappa $, with
\begin{align}
\label{eq:ir_reg}
\Delta S_\kappa[\varphi] = \frac{1}{2}\int_{x,x'} \varphi_a(x) R_{\kappa}(x,x') \varphi_a(x'),
\end{align}
where the infrared regulator $R_{\kappa}$ acts as a large mass term for (quantum) fluctuations on sizes larger than $1/\kappa$ and essentially vanishes for short wavelength modes, thereby suppressing the contribution from the former to the path integral. The idea is then to progressively integrate out the potentially dangerous infrared fluctuations by continuously  changing the scale $\kappa$ from a ultraviolet scale\footnote{The scale $\Lambda$ can be taken to infinity for renormalizable theories.} $\kappa=\Lambda$, where the (bare) theory is defined, down to $\kappa=0$. One defines a regularized effective action $\Gamma_\kappa$ which interpolates between the bare action, $\Gamma_{\kappa=\Lambda}=S$ and the standard effective action, i.e., the generating functional of one-particle-irreducible vertex functions, $\Gamma_{\kappa=0}=\Gamma$. It satisfies an exact (functional) flow equation \cite{Wetterich:1992yh,Morris:1993qb} 
\eq{
\dot \Gamma_\kappa[\phi] = \frac{1}{2}\Tr \left\{\dot R_\kappa \ast G_\kappa\right\},\label{eq:wetterich}
}
where a dot denotes a derivative with respect to the RG time $\tau=\ln\kappa$. The functional trace $\Tr$ and convolution product $\ast$ refer to a given integration measure. We use the covariant measure $\int_x$ defined above. Accordingly, we define the covariant two-point vertex function
\eq{\label{eq:Gamma2}\Gamma_{\kappa,ab}^{(2)}[\phi](x,y)= \frac{\delta^2_c\Gamma_\kappa[\phi]}{\delta\phi_a(x)\delta\phi_b(y)},}
where we have defined the covariant functional derivative as
 \eq{\frac{\delta_c}{\delta\phi(x)}=\frac{1}{\sqrt{-g(x)}}\frac{\delta}{\delta\phi(x)}.}
The two-point function \eqn{eq:Gamma2} relates to the exact propagator $G_\kappa$ of the regulated theory as
\eq{G_\kappa=i\left(\Gamma_\kappa^{(2)} + R_\kappa\right)^{-1},}
where the inversion refers to the convolution product $\ast$. 

There are two technical points to be emphasized here. The first one is that we are primarily interested in computing the correlation functions of the theory in a given initial state  specified in the infinite past; see below. Such an initial-value problem is most conveniently formulated in the {\it in-in}, or closed-time-path formalism \cite{Schwinger:1960qe,Berges:2004yj}, where time integrations are to be understood along Schwinger's closed time contour ${\cal C}$, which goes from infinite past to infinite future and back. In the present context, this amounts, e.g., to the replacements $\int d\eta\to\int_{\cal C}d\eta$ and $\delta(\eta-\eta')\to\delta_{\cal C}(\eta-\eta')$; see Ref.~\cite{Parentani:2012tx} for details.\footnote{Discussions of NPRG methods for nonequilibrium systems can be found in Refs.~\cite{Gasenzer:2008zz,Canet:2011wf,Berges:2012ty}.} 

The second, related point concerns the issue of the de Sitter isometries. In general, one chooses a regulator function $R_\kappa$ which respects as many symmetries of the problem at hand as possible, in order to ensure that the resulting RG flow respects the latter down to $\kappa=0$. In the present case, it is not clear how to construct a proper regulator which respects all the de Sitter isometries.\footnote{This is, in fact, a general issue for space-times with Lorentzian signature, in relation with causality; see, e.g., \cite{Canet:2011wf}.} 
Here, we follow Refs.~\cite{Kaya:2013bga,Serreau:2013eoa,Guilleux:2015pma} and choose an infrared regulator of the form 
\begin{align}
 R_\kappa(x,x')&=-\frac{\delta_{\cal C}(\eta-\eta')}{a^D(\eta)}\int\frac{d^dK}{(2\pi)^d}e^{i{\bf K}\cdot({\bf X}-{\bf X}')} \hat R_\kappa(-K\eta)\nn
 \label{eq:regreg}
 &=-\delta_{\cal C}(t-t')\int\frac{d^dp}{(2\pi)^d}e^{i{\bf p}\cdot({\bf x}-{\bf x}')} \hat R_\kappa(p),
\end{align}
which preserves a large subset of de Sitter isometries\footnote{In the expanding Poincar\'e patch this corresponds to the subgroup of space and time translations on the hyperboloid  and spatial rotations. The former are generated by the spacelike and timelike Killing vector fields ${K}^i_{ s}=\partial_{X^i}|_t$ and $K_{ t}=\partial_t|_{x^i}=\partial_t|_{X^i}-H{X}^iK_s^i$, where $t$ is the cosmological time whereas $x^i$ and $X^i$, with $i=1,\ldots,d$, are the physical and comoving spatial coordinates, respectively \cite{Eling:2006xg,Busch:2012ne}. Together with the generators of spatial rotations $J^{ij}=x^iK_s^j-x^jK_s^i$, the subgroup algebra is 
\eq{[K_{ t},K_{ s}^i]&=H{K}_{ s}^i,\\
[K_s^i,J^{jk}]&=\delta^{ij}K_s^k-\delta^{ik}K_s^j,\\
[J^{ij},J^{kl}]&=-\delta^{ik}J^{jl}+\delta^{il}J^{jk}+\delta^{jk}J^{il}-\delta^{jl}J^{ik},
}
and all the other commutators vanish.} \cite{Busch:2012ne,Adamek:2013vw,Parentani:2012tx}. Here, in the second line, we have introduced physical coordinates and momentum variables, ${\bf x}=a(\eta){\bf X}$ and ${\bf p}={\bf K}/a(\eta)$. The important point here is that we regulate physical momenta (as opposed to comoving ones). Interestingly, this induces an effective regulation of the time variable because of the way momentum and time are tight together by the gravitational redshift.

As already mentioned, the flow equation \eqn{eq:wetterich} is exact. However, such a nonlinear functional partial differential equation is in general not solvable exactly and one has to use approximations. The simplest local potential {\it Ansatz} (LPA) has been discussed at length in previous works \cite{Kaya:2013bga,Serreau:2013eoa,Guilleux:2015pma} and produces interesting physical results. The purpose of the present work is to explore the possible formulation and applications of approximations beyond the LPA, in particular, the derivative expansion that we now discuss.

\subsection{Derivative expansion and LPA'}\label{sec:derivexp}

The derivative expansion has been widely used in statistical physics applications of the NPRG \cite{Morris:1994ie,Delamotte:2007pf}. It is a systematic expansion in powers of derivatives of the field, which aims at capturing the dynamics of long wavelength excitations, relevant, e.g., for computing critical exponents. It appears that this very idea is not completely straightforward in a general curved space-time because of possible couplings to the Riemann tensor and its contractions. In general, one expects the typical variation of the field to be related to those of the curvature and a sensible derivative expansion is likely to count gradients of both quantities on an equal footing.\footnote{ One could instead count derivatives of the metric tensor (see, e.g., Ref.~\cite{Shapiro:2015ova}) but this seems to us unphysical since the latter can be made arbitrarily large by (in)appropriate choices of coordinates.}  At first nontrivial order in gradients, this includes the standard term $\nabla_{\!\mu}\phi \nabla^\mu \phi$, but also terms of the form\footnote{A term $\nabla_\mu R^{\mu\nu}\nabla_\nu\phi$ is also possible but not independent since $\nabla_\mu R^{\mu\nu}=\frac{1}{2}\nabla^\nu R$.} $\nabla_{\!\mu} R \nabla^\mu \phi$, $R^{\mu \nu} \nabla_\mu\phi \nabla_\nu \phi$, and $R^{\mu \nu} \nabla_\mu R \nabla_\nu \phi$. Furthermore, all the coefficients of the gradient terms (including the zeroth-order local potential term) should be seen as functions of the field $\phi$ and all scalars made of the Riemann tensor, such as $R$, $R_{\mu\nu}R^{\mu\nu}$, etc. For instance, keeping the full curvaturedependence in the potential is crucial in order to correctly capture the physics of superhorizon modes. In de Sitter space, this leads to nontrivial effects such as dynamical mass generation with $m^2\propto R$ \cite{Starobinsky:1994bd}. Finally, to make matters even more intricate, we mention that the non-commuting covariant derivatives make the separation between different orders in gradient ambiguous. For instance, the apparently fourth-order term
\eq{\nabla_\mu \nabla_\nu \nabla^\mu \nabla^\nu \phi=\Box^2 \phi + \frac{1}{2} \nabla_{\!\mu} R \nabla^\mu \phi + R_{\mu \nu} \nabla_\mu \nabla_\nu \phi,}
where $\Box= \nabla_\mu \nabla^\mu $ is the Laplace-Beltrami operator, contains second-order contributions. In brief, a derivative expansion is far from being trivial (if implementable at all) in a general space-time.

Of course, matters simplify in space-times with a large number of symmetries. In particular, in the maximally symmetric, de Sitter space, the Riemann tensor is covariantly conserved, $\nabla_\mu R_{\alpha\beta\gamma\delta}=0$, and all scalar contractions are constant, e.g., $R=d(d+1)H^2$, $R_{\mu\nu}R^{\mu\nu}=d^2(d+1)H^4$, etc. One can thus devise a systematic expansion in powers of, e.g., $\Box\phi$, with coefficients depending on the field only.\footnote{For instance, one has $\nabla_\mu \nabla_\nu \nabla^\mu \nabla^\nu \phi = \Box^2 \phi + d H^2 \Box \phi$.}  For a theory with a single scalar field, one writes 
\eq{
\Gamma_\kappa [\phi] &= -\int_x \left\{V_\kappa(\phi) + \frac{Z_\kappa(\phi)}{2}\partial_\mu\phi\partial^\mu \phi + {\cal O}\left(\partial^4\right) \right\}. \label{eq:ansatz}
}
Yet, this is not the end of the story...

The flow of the local potential $V_\kappa$ can be obtained by writing \Eqn{eq:wetterich} with the {\it Ansatz} \eqn{eq:ansatz} for constant field configuration. To derive the flow equation for the kinetic term $Z_\kappa$, we consider the two-point vertex function \eqn{eq:Gamma2} for a constant field configuration, namely, at second order in gradients,
\eq{\label{eq:G2der}
 \Gamma_{\kappa}^{(2)}(x,x') =\Big[ -V_\kappa ''(\phi) +Z_\kappa(\phi) \Box+{\cal O}\left(\Box^2\right) \Big] { \delta(x, x')}. 
}
where we denote the covariant Dirac distribution on the time contour as
\eq{ \delta(x, x') = \delta_c(\eta,\eta')\delta^{(d)}({\bf X}-{\bf X}'),}
with
\eq{\delta_c(\eta,\eta') = {\delta_{\cal C}(\eta - \eta')\over a^D(\eta)}.}
In Minkowski space-time, we could now exploit the translation invariance in both the spatial and the temporal directions, valid for constant fields, and diagonalize the operator $\Box$ by going to $D$-dimensional Fourier space, i.e., $\Box\to-k^2$ and  $\Gamma_\kappa^{(2)}(x-x')\to \Gamma_\kappa^{(2)}(k^2)$. The RG flows of the various renormalization factors of the derivative expansion can then be obtained as the coefficients the momentum expansion of $\dot \Gamma_\kappa^{(2)}(k^2)$ around $k^2=0$. For instance, one would have $\dot Z_\kappa(\phi)=-\partial_{k^2}\dot \Gamma_\kappa^{(2)}(k^2)|_{k^2=0}$.

Clearly, this step is not as simple in a general curved space-time because of the lack of space and/or time translation symmetries. But even in the maximally symmetric de Sitter space, which possesses translational invariances both in the (cosmological) time and in the (comoving) spatial directions, the problem remains tricky because the corresponding Killing fields do not commute and thus cannot be diagonalized simultaneously.

Exploiting spatial translation and rotation invariance in comoving coordinates and introducing the Fourier transform 
\eq{\label{eq:Fourier}\Gamma_{\kappa}^{(2)}(x,x')&=\int \frac{d^dK}{(2\pi)^d}\,e^{i{\bf K}\cdot({\bf X}-{\bf X}')}\,\Gamma_\kappa ^{(2)}(\eta, \eta', K),}
we have
\eq{\label{eq:14}
\Gamma_\kappa ^{(2)}(\eta, \eta', K) = \Big [{-}V_\kappa''(\phi) \!+\! Z_\kappa(\phi) \Box_{K}  {+}{\cal O}\left(\Box_K^2\right) \!\Big]  \delta_c(\eta,\eta'),
}
where we have noted $\Box_{K} = -\left(\eta\partial_{\eta}\right)^2 +d\eta \partial_{\eta} -K^2 \eta^2$ the Laplacian at fixed $K$. At first sight, an easy way to extract the flow of the coefficients of the derivative expansion seems to consider an expansion in $K$ around $K=0$, similar to the flat-space case. This is, however, too naive, again because of the noncommutation of the generators of space and time translations or, equivalently, because of the redshift of physical momenta encoded in the term $K^2\eta^2$. Indeed, it is easy to check that higher-order derivative operators $\Box_K^n$, with $n\ge1$, all contribute a $K^2\eta^2$ term. For instance, 
\eq{\Box_K^2=\Box_{K=0}^2 -2K^2 \eta^2\Box_{K=0}-(2d-4)K^2\eta^2+K^4\eta^4.} 
More generally, all operators $\Box_K^n$ contain terms $(K^2\eta^2)^i$ with $i\le n$. It follows that the expansion of $\Gamma_\kappa ^{(2)}(\eta, \eta', K)$ in powers of $K^2\eta^2$ does not coincide with the derivative expansion: each coefficient of the former actually mixes an infinite number of terms from the latter. 

The bottom line of the above discussion is that one should expand the function $\Gamma_{\kappa}^{(2)}(x,x')$ on a basis of eigenfunctions of the operator $\Box$, which is clearly not the case of the spatial plane waves underlying the Fourier decomposition \eqn{eq:Fourier}. As a practical constraint, it is also desirable that the relevant set of eigenfunctions be simple enough so that one can perform actual calculations. In this spirit, we propose to extract the coefficients of the derivative expansion by considering the Fourier representation \eqn{eq:Fourier} at $K=0$. The relevant operator is then $\Box_{K=0}= -\partial_t^2-d\partial_t$, with $\partial_t=-\eta\partial_\eta$, which is diagonalized by the functions $e^{-i\omega t}\propto \eta^{i\omega}$, with $\omega \in \mathbb{C}$. We have
\eq{
\Box_{K=0} \,\eta^{i\omega}  = \alpha_\omega\eta^{i\omega}. \label{eq:time_diag}
}
with $\alpha_\omega=\omega^2+ id\omega$.
Accordingly, we introduce the following Fourier-like transform along the time contour:
\eq{
\Gamma_{\kappa} ^{(2)}(\omega) = \int_{\cal C} d\eta' a^D(\eta') \pfrac{\eta'}{\eta}{\!\!i\omega} \Gamma_{\kappa} ^{(2)}(\eta, \eta',K=0), \label{eq:laplace}
}
and we shall perform an expansion in powers of $\alpha_\omega$ in order to isolate the coefficients of the derivative expansion. Indeed, we have 
\eq{\label{eq:local}
\Gamma_{\kappa}^{(2)}(\omega) = - V_{\kappa}''(\phi) + Z_\kappa(\phi)  \alpha_\omega + O(\alpha_\omega^2),
}
from which we get the flows of the desired functions, e.g., 
\eq{\label{eq:floVsec}\dot V_{\kappa}''(\phi)&=-\dot\Gamma_{\kappa}^{(2)}(\omega=0),\\
\label{eq:floZphi}\dot Z_\kappa(\phi)&=\left.\frac{\partial\dot\GammaÄ_{\kappa}^{(2)}(\omega)}{\partial\alpha_\omega}\right|_{\omega=0}.
} 

Some comments are in order:

\begin{itemize}[leftmargin=*]

\item First, the transform \eqn{eq:laplace} is time independent due to the de Sitter symmetries. Using the exact scaling relation \cite{Parentani:2012tx}
\eq{
\Gamma_\kappa^{(2)}( \alpha \eta, \alpha \eta',K/\alpha) =\alpha^{d}\, \Gamma_\kappa^{(2)}(\eta, \eta', K),
}
with $\alpha\in\mathds{R}$, in \eqn{eq:laplace}, one gets
\eq{
\Gamma_{\kappa} ^{(2)}(\omega) =\! \int_{\cal C} d\eta' a^D(\eta') \!\pfrac{\eta'}{\alpha \eta}{\!\!i\omega} \!\Gamma_{\kappa} ^{(2)}(\alpha \eta, \eta',K=0)
}
after the change of variable $\eta' \to \eta'/\alpha$, hence, the announced result.

\item Second, the transform \eqn{eq:laplace} on the closed time contour can be expressed as a standard Fourier transform along the real time axis in terms of the cosmological time and properly rescaled quantities. Indeed, using the identity
$(\eta'/\eta)^{i\omega} = e^{i\omega(t-t')}$ and introducing the  function\footnote{ It is this rescaled function which depends only on $t-t'$ in the limit of subhorizon physical momenta $K/a(\eta), K/a(\eta')\gg1$ and time difference $|t-t'|\ll1$. Indeed, in this limit, the rescaled propagator $\bar G_\kappa^{(2)}(t,t',K)=[a(\eta)a(\eta')]^{d/2}G_\kappa^{(2)}(\eta,\eta',K)$ tends to the Minkowski propagator \cite{Serreau:2013psa}, which only depends on time through the difference $t-t'$. It is easy to show that this rescaled propagator is related to the vertex function \eqn{eq:bargamma} through $\int_{\cal C}dt''\bar\Gamma_\kappa^{(2)}(t,t'',K)\bar G_\kappa^{(2)}(t'',t',K)=i\delta_{\cal C}(t-t')$, where we have used the fact that the regulator function vanishes for momenta $K/a\gg\kappa$. It follows that $\bar\Gamma_\kappa^{(2)}(t,t',K)$ only depends on time through the time difference $t-t'$ in the subhorizon limit.} 
\eq{\label{eq:bargamma}
\bar\Gamma_\kappa^{(2)}(t,t',K)=[a(\eta)a(\eta')]^{d/2}\Gamma_\kappa^{(2)}(\eta,\eta',K),
}
we have
\eq{
\Gamma_{\kappa} ^{(2)}(\omega) = \int_{\cal C} dt' e^{\left(i\omega-{d\over2}\right)(t-t')}\bar \Gamma_{\kappa} ^{(2)}(t, t',K=0),
}
Moreover, introducing the statistical ($F$) and spectral $\rho$ components of any nonlocal\footnote{Local terms, proportional to $\delta_{\cal C}(t-t')$ and its derivatives, can easily be treated as in \Eqn{eq:local}.} two-point function on the contour $A(t, t')$ as \cite{Berges:2004yj}
\eq{
 A(t, t')=A_F(t, t')-\frac{i}{2}{\rm sign}_{\cal C}(t-t')A_\rho(t, t'),
}
one has 
\eq{\label{eq:retardedtransform}
i\!\int_{\cal C}dt'A(t,t')=\int_{-\infty}^t\! \!dt' A_\rho(t,t')=\int_{-\infty}^{+\infty} \!\!dt'A_R(t,t'),
} where, in the second equality, we further introduced the retarded two-point function 
\eq{
A_R(t,t')=\theta(t-t')A_\rho(t,t').
}
We see that the transform \eqn{eq:laplace} is given by the Fourier transform (in cosmological time) of the retarded component of the two-point function $\bar\Gamma_\kappa^{(2)}(t,t',K=0)$ evaluated at a complex frequency:
\eq{\label{eq:FTR}
\Gamma_{\kappa}^{(2)}(\omega) = -i\bar\Gamma_{R,\kappa}^{(2)}\left(\omega+i{d/2},K=0\right)
} 

\item The approach described above reduces to straightforward prescriptions in the flat space-time, Minkowski limit $H\to0$. Making factors $H$ explicit, we replace $\omega$ by $\omega/H$ and $\square$ by $H^2\square$ in the above discussions such that $\alpha_\omega=\omega^2+idH\omega\to\omega^2$ in Eqs.~\eqn{eq:time_diag} and \eqn{eq:floZphi}. Similarly, $\omega+idH/2\to\omega$ in \Eqn{eq:FTR}, so that
\eq{
\Gamma_{\kappa} ^{(2)}(\omega) \to-i\bar\Gamma_{R,\kappa}^{(2)}\left(\omega,K=0\right) , \label{eq:laplaceflatlimit}
}
with $\bar\Gamma_{R,\kappa}^{(2)}\left(\omega,K\right)$ the $D$-dimensional Fourier transform of the function $\bar\Gamma_\kappa^{(2)}(t,t',K)$ introduced in \eqn{eq:bargamma}. Note that, in the limit $H\to0$, the latter only depends on times through the difference $t-t'$ and $\Gamma_{\kappa} ^{(2)}(\omega,K)$ is time-independent for any $K$, reflecting the fact that the generators of space and time translations now commute.

\item A drawback of our approach is that it implicitly assumes that the regulator, hence the flow, respects the de Sitter symmetries and, in particular, the symmetric role of timelike and spacelike gradients in the operator $\Box$. However, as we have emphasized previously, our regulator only respects a subgroup of the de Sitter isometries and treats time and space differently. This implies that the various derivative terms compatible with the symmetries (see, e.g., Ref.~\cite{Busch:2012ne}) of the regulated theory should receive {\it a priori} different renormalization factors. For instance, at second order, the time and space gradient terms would receive independent renormalization factors: $ Z^{\rm t}_\kappa(\phi) \Box_{K=0}-Z^{\rm s}_\kappa(\phi)K^2\eta^2$. If the method outlined above allows one to consistently extract the flow of $Z_\kappa^{\rm t}$, the discussion below \Eqn{eq:14} shows that it is not clear how to unambiguously extract that of $Z^{\rm s}_\kappa$. Nevertheless, the whole approach---with the present regulator---makes sense only if the explicit symmetry breaking induced by the regulator is small, in which case it is enough to compute the flow of $Z_\kappa^{\rm t}$ following the above procedure. We test these ideas in Appendix~\ref{appsec:Mink} in the case of Minkowski space, where we can explicitly compute both $\dot Z_\kappa^{\rm s}$ and $\dot Z_\kappa^{\rm s}$. Of course, a possible way out would be to construct a fully de Sitter invariant regulator. We postpone a detailed study of these issues to a later work and, from now on, we simply assume that such symmetry breaking effects can be neglected. 
\end{itemize}

In the following, as a proof of principle of the feasibility of the above program, we consider a somewhat simplified version of the derivative expansion, where one neglects the field dependence of the field renormalization factor $Z_\kappa(\phi)\to Z_\kappa$. This so-called local potential approximation prime, or LPA', corresponds to the {\it Ansatz}
\eq{
\Gamma_\kappa^{\rm LPA'} [\phi] = -\int_x \left\{V_\kappa(\phi) + \frac{Z_\kappa}{2}\partial_\mu\phi\partial^\mu \phi \right\}. \label{eq:LPAprime}
}
This has already been considered in Ref.~\cite{Serreau:2013eoa} where, however, the flow of $Z_\kappa$ was not explicitly computed. It is the purpose of the following section to compute the latter and its influence on the flow of the local potential. Because the left-hand side of \Eqn{eq:floZphi} does not depend on $\phi$ anymore whereas the right-hand side does (this is part of the inconsistencies of any {\it Ansatz}), we have to specify a value of $\phi$ where to evaluate it. We follow the standard practice in this context and choose the minimum of the running potential $V_\kappa(\phi)$. We thus define the running anomalous dimension as
\eq{
\eta_\kappa = -\frac{\dot Z_\kappa}{Z_\kappa}= -\frac{1}{Z_\kappa} \frac{\partial \dot \Gamma_{\kappa} ^{(2)}(\omega)}{\partial\alpha_\omega} \bigg |_{\omega=0,{\rm min}}. \label{eq:eta_precsription}
}

\section{Anomalous dimension for a single field}
\label{sec:N1}

\subsection{Flow equation}

For the present purposes, it is convenient to rewrite the flow equation \eqn{eq:wetterich} in the form \cite{Berges:2000ew}
\eq{\dot \Gamma_\kappa[\phi] &= \frac{i}{2} \tilde\partial_\tau \Tr \,{\rm Ln} \left(\Gamma_\kappa^{(2)}+R_\kappa\right),}
where the derivative $\tilde\partial_\tau$ only acts on the explicit regulator dependence on the right-hand side. This form of the equation makes it particularly simple to obtain the flow of the two-point vertex function:
\eq{\label{eq:flogtwo}
&\dot {\Gamma}_{\kappa}^{(2)}(x,y) = \frac{1}{2} \tilde\partial_\tau \! \int_{a,b} \Gamma_{\kappa}^{(4)}(x,y,a,b) G_{\kappa }(b,a)  \nn
& \qquad+i  \int_{a,b,c,d} \Gamma_{\kappa}^{(3)}(x,a,b)G_{\kappa}(a,c)   G_{\kappa}(b,d)  \Gamma_{\kappa}^{(3)}(c,d,y),
}
where
\eq{\Gamma_\kappa^{(n)}(x_1,\ldots,x_n)=\frac{\delta_c^n\Gamma_\kappa}{\delta\phi(x_1)\ldots\delta\phi(x_n)}.}
We must compute the three- and four-point vertex functions using our preferred {\it Ansatz}. This is where the LPA'  {\it Ansatz} \eqn{eq:LPAprime} greatly simplifies matters. The derivative term being quadratic in the field, does not contribute to the three- and higher-point vertices. We have
\eq{
\Gamma_{\kappa}^{(3)}(x,y,z) &=  - V_{\kappa}^{(3)}(\phi) \delta(x,y) \delta(y,z), \\
\Gamma_{\kappa}^{(4)}(w,x,y,z) &= - V_{\kappa}^{(4)}(\phi) \delta(x,y) \delta(y,z) \delta(w,x),
}
and the flow equation \eqn{eq:flogtwo} becomes
\begin{equation}
\dot {\Gamma}_{\kappa}^{(2)}(x,y)\!= \!\frac{1}{2} { \tilde \partial_\tau}  \!\left\{ \!- V_{\kappa}^{(4)}G_{\kappa}(x,x) \delta(x,y) \!+\!i {V_{\kappa}^{(3)}}^2 G^2_{\kappa}(x,y)\!\right\}\!.
\end{equation}
After exploiting spatial homogeneity and isotropy, this yields, in comoving spatial Fourier space, 
\eq{
\dot \Gamma_\kappa^{(2)}(\eta,\eta',K) = \frac{1}{2} { \tilde \partial_\tau} \!\int_Q \bigg\{ \!\!-V_\kappa^{(4)}   G_\kappa(\eta,\eta,Q) \delta_c(\eta,\eta')  &\label{eq:Gflow}\nonumber\\
 +  i {V_\kappa^{(3)}}^2   G_\kappa(\eta,\eta',Q)  G_\kappa(\eta,\eta',L) &\bigg\}
}
with $L=|{\bf K}-{\bf Q}|$ and $\int_Q=\int d^dQ/(2\pi)^d$. Finally, we take the transform \eqn{eq:laplace} and we use the physical momentum representation of correlators \cite{Parentani:2012tx}
\eq{\label{eq:prep}
G_\kappa(\eta,\eta',K) = \frac{(\eta \eta')^{\frac{d-1}{2}}}{K} \hat G_{ \kappa}(p, p'),
}
with $p=-K\eta$ and $p'=-K\eta'$. This exact scaling relation is a consequence of the de Sitter isometries (in fact, of the subgroup mentioned previously) which precisely states how physical momenta get correlated by the gravitational redshift. It allows one to scale out the comoving momentum $K$ and to deal the time evolution for a physical momentum evolution. Accordingly, one introduces a closed contour $\hat{\cal C}$ in momentum; see Ref.~\cite{Parentani:2012tx} for details.
It is straightforward to show that the transform \eqn{eq:laplace} writes
\eq{\label{eq:flowomega}
\dot \Gamma_\kappa^{(2)}(\omega) &= -\frac{\Omega_d}{(2\pi)^d} { \tilde \partial_\tau}\!\int_0^{{\infty}} dp \ p^{d-2} \bigg\{  \frac{V_\kappa^{(4)} }{2} \hat F_{ \kappa}(p,p)  \nn
 &+ {V_\kappa^{(3)}}^2\!  \int_p^{\infty} \frac{dp'}{p'^2} \pfrac{p'}{p}{\!\!i\omega} \!\hat F_{ \kappa}(p,p') \hat \rho_{ \kappa}(p',p) \bigg\},
}
where $\hat F_\kappa$ and $\hat \rho_\kappa$ are respectively the statistical and spectral correlators defined as 
\eq{
\hat G_\kappa (p,p') = \hat F_\kappa(p,p') - \frac{i}{2}{\rm sign}_{\hat {\cal C}}(p-p') \hat \rho_\kappa(p,p').
}
It is useful to note the symmetry properties $\hat F_\kappa(p,p')=\hat F_\kappa(p',p)$ and $\hat\rho_\kappa(p,p')=-\hat\rho_\kappa(p',p)$.

To compute the derivative ${ \tilde \partial_\tau}$ acting on these propagators, we consider the variation $R_\kappa \to R_\kappa + \delta R_\kappa$ in the relation $(\hat \Gamma^{(2)}_\kappa+\hat R_\kappa)\ast\hat  G_\kappa = i$, which yields
\eq{
{ \tilde \partial_\tau}\hat G_\kappa = i\hat G_\kappa \ast \dot{\hat  R}_\kappa\ast \hat G_\kappa.
}
By identifying the statistical and spectral parts on both sides, we get
 \begin{align}
{ \tilde \partial_\tau} \hat F_\kappa(p,p') = &- \int_p^{{\infty}} \!\!dr\, \hat \rho_\kappa(p,r) \frac{\dot{\hat R}_\kappa(r)}{r^2}\hat F_\kappa(r,p')  \\
&+ \int_{p'}^{{\infty}} \!\!dr \,\hat F_\kappa(p,r) \frac{\dot{\hat R}_\kappa(r)}{r^2}\hat \rho_\kappa(r,p')  \nonumber \\
{ \tilde \partial_\tau} \hat \rho_\kappa(p,p') =& - \int_p^{{p'}}  \!\!dr \, \hat \rho_\kappa(p,r) \frac{\dot{\hat R}_\kappa(r)}{r^2} \hat \rho_\kappa(r,p') 
 \end{align}
To explicitly compute the flow \eqn{eq:Gflow} we need to specify the regulator function. We choose \cite{Litim:2001up} 
\eq{\hat R_\kappa(p) = Z_\kappa(\kappa^2 - p^2)\theta(\kappa^2 - p^2),\label{eq:regulator}} 
for which $\dot{ \hat R}_\kappa(p)=Z_\kappa\left[(2-\eta_\kappa)\kappa^2 + \eta_\kappa p^2\right]\theta(\kappa^2 - p^2)$.This simple form allows one to perform some of the integrals analytically. After some algebra, the resulting flow can be written as
\eq{
\dot \Gamma_\kappa ^{(2)}(\omega) = \frac{Z_\kappa \Omega_d}{(2\pi)^d} \left (  V_\kappa^{(4)} J_\kappa  + {V_\kappa^{(3)}}^2 \sum_{n=0}^4 I_\kappa^{(n)} \right ) ,\label{eq:gamma_dot}
}
where we have left the field dependence on both sides implicit for simplicity and where we defined the integrals
\eq{\label{eq:Jdef}
J_\kappa &=  \int_0^{\mathrlap{\kappa}}dp \! \int_p^{\mathrlap{\kappa}} dr\,\, p^2\!A(p,p,r)\,\hat \rho_\kappa(p,r) \hat F_\kappa(r,p)
}
and
\eq{
I_\kappa^{(0)} &=  \int_0^{\mathrlap{\kappa}}dp\! \int_p^{\mathrlap{\kappa}} dq \!\int_p^{\mathrlap{\kappa}} dr \,A(p,q,r) \hat \rho_\kappa(p,r)\hat F_\kappa(r,q) \hat \rho_\kappa(q,p) ,\label{eq:I_0} \\
I_\kappa^{(1)} &=  \int_0^{\mathrlap{\kappa}}dp\! \int_\kappa^{\mathrlap{\infty}} dq \!\int_p^{\mathrlap{\kappa}} dr\, A(p,q,r) \hat \rho_\kappa(p,r)  \hat F_\kappa(r,q) \hat \rho_\kappa(q,p) ,\label{eq:I_1}\\
I_\kappa^{(2)} &= - \!\int_0^{\mathrlap{\kappa}}dp\! \int_p^{\mathrlap{\kappa}} dq \!\int_{q}^{\mathrlap{\kappa}} dr\, A(p,q,r) \hat F_\kappa(p,r)  \hat \rho_\kappa(r,q) \hat \rho_\kappa(q,p) ,\label{eq:I_2}\\
I_\kappa^{(3)} &=  - \!\int_0^{\mathrlap{\kappa}}dp \!\int_p^{\mathrlap{\kappa}} dq \!\int_p^{\mathrlap{q}} dr\, A(p,q,r) \hat \rho_\kappa(p,r)  \hat \rho_\kappa(r,q) \hat F_\kappa(q,p) ,\label{eq:I_3}\\
I_\kappa^{(4)} &= - \!\int_0^{\mathrlap{\kappa}}dp\! \int_\kappa^{\mathrlap{\infty}} dq \!\int_p^{\mathrlap{\kappa}} dr \,A(p,q,r) \hat \rho_\kappa(p,r)  \hat \rho_\kappa(r,q) \hat F_\kappa(q,p), \label{eq:I_4}
}
with the integration measure
\eq{
A(p,q,r)= p^d \left ( \dfrac{q}{p} \right)^{\!\!i\omega} \frac{(2-\eta_\kappa)\kappa^2 + \eta_\kappa r^2}{(pqr)^2}.
}
Here, we have split the contributions with all momenta below the RG scale $\kappa$ from those which involve modes above this scale. Specifically, the contributions $I_\kappa^{(1)}$ and $I_\kappa^{(4)}$ involve modes $q>\kappa$ which, as a result of the gravitational redshift, have nontrivial correlations with modes $p,r<\kappa$. As we shall see, these contributions vanish in the Minkowski limit where the redshift is absent. They also give subdominant contributions in the infrared limit.

Another consequence of the gravitational redshift is the fact that a cutoff scale on physical momenta effectively restricts the range of time integration. This is visible on Eqs.~\eqn{eq:Jdef}--\eqn{eq:I_4} where the variables $r$ and $q$ originally arise from time integrations. To illustrate this further, consider the change of variables $(p,r)\to(K,t')$, with $p=Ke^{-Ht}$ and $r=Ke^{-Ht'}$ for a given $t$ in the integral \eqn{eq:Jdef}, where we have made the dimensionful factors $H$ explicit.\footnote{Note that the mass dimension of $J_\kappa$ is $d-1$. The various powers of momenta in the definition \eqn{eq:Jdef} contribute for $d$. Hence, one must include a factor $1/H$ on the right-hand side of \Eqn{eq:Jdef}. A similar analysis shows that one must include a factor $1/H^2$ in the defining equations of $I_\kappa^{(n)}$, whose mass dimension is $d-3$.} Introducing 
\eq{\bar G_\kappa(t,t',K)=[a(\eta)a(\eta')]^{d/2}G_\kappa(\eta,\eta',K),}
this yields, taking the case $\eta_\kappa=0$ for illustration,
\eq{\label{eq:expression}J_\kappa=\frac{2\kappa^2}{\bar a^d(t)}\int_0^{{ \kappa\bar a(t)}}\!\!dK K^{d-1} \!\! \int_{t_K }^{\mathrlap{t}} \!dt'\, \bar \rho_\kappa(t,t',K)\bar F_\kappa(t',t,K),}
where $\bar a(t)=a(\eta)=e^{Ht}$. The time integration is bounded by the time\footnote{Note that $t_K\le t$ for $K\le \kappa \bar a(t)$ in \Eqn{eq:expression}.} $t_K =-{1\over H} \ln (\kappa/ K)\le t$ at which the physical momentum $K/\bar a(t)$ crosses the running scale $\kappa$. The expression \eqn{eq:expression} also makes clear how this effect of the gravitational redshift disappears in the flat-space limit $H\to0$, where $t_K\to-\infty$. In this case, one gets
\eq{J_\kappa\to2\kappa^2\int_0^{\kappa}dK K^{d-1}  \int_{-\infty }^{\mathrlap{t}} dt' \bar \rho_\kappa(t,t',K)\bar F_\kappa(t',t,K),}
which is, indeed, the result one obtains by applying the present formalism directly in Minkowski space \cite{Guilleux:2015pma,Maximethesis}.
This discussion easily generalizes to the integrals $I_\kappa^{(0,2,3)}$ with the further change of variable $q=Ke^{-Ht''}$. In that case, the factor $(q/p)^{i\omega\over H}= e^{i\omega(t-t'')}$; see \Eqn{eq:laplaceflatlimit}. As for the integrals $I_\kappa^{(1,4)}$, which involve modes $q\ge\kappa$, the above analysis yields a time integral $\int_{-\infty}^{t_K}dt''$, which guarantees that $I_\kappa^{(1,4)}\to 0$ in the flat-space limit. 

The flow equation \eqn{eq:gamma_dot} is quite complicated and cannot be written in a simple form in general. However, it greatly simplifies in two opposite limits that we now discuss, where all dimensionful scales are either large or small in units of the curvature. The first one corresponds to the Minkowski limit discussed above while the second is the one  of prime interest to us, where curvature effects become important. At this point, it is worth emphasizing that, in the LPA', only the contribution on the second line of \Eqn{eq:flowomega} depends on the variable $\omega$, from which it follows that $\eta_\kappa\propto V_\kappa^{(3)\,2}|_{\rm min}$. Because our prescription is to evaluate \Eqn{eq:eta_precsription} at the minimum of the potential,  $\eta_\kappa$ is identically zero in the symmetric phase. As a consequence, we will be interested in cases where the potential presents a spontaneous symmetry breaking shape along the RG flow. In particular we shall study the possible effect of the running anomalous dimension $\eta_\kappa$ on the phenomenon of symmetry restoration in the deep infrared regime.

\subsection{Heavy UV regime: Minkowski flow}

The integrals \eqn{eq:Jdef}--\eqn{eq:I_4} can be computed analytically in the limit where both the RG scale and the curvature of the running potential are large in units of $H$, which is equivalent to sending $H\to0$. We shall not reproduce this instructive but cumbersome calculation here. We refer the reader to Ref.~\cite{Maximethesis} for details. Instead, we shall compare with a calculation of the LPA' flow equations directly in Minkowski space, detailed in Appendix~\ref{appsec:Mink}. The $H\to0$ limit of \Eqn{eq:gamma_dot} yields 
\eq{\label{eq:Minko}
\dot \Gamma_\kappa^{(2)}(\omega) &=\frac{v_d\kappa^{d+2}}{2M_\kappa^3} \bigg(1-\frac{\eta_\kappa}{d+2} \bigg) \nn
&\times\left( \frac{ V^{(4)}}{Z_\kappa}+ \frac{\mbox{${ {V}}_\kappa^{(3)}$}^2}{Z_\kappa^2} \frac{2\omega^2-24 M_\kappa^2}{(\omega^2-4M_\kappa^2)^2} \right),
}
where $v_d =  \Omega_d /[2 d(2\pi)^d]$ and $M_\kappa^2=\kappa^2+V_\kappa''/Z_\kappa$ is the regulated potential curvature. Setting $\omega=0$ gives the flow of the curvature of the potential [see \Eqn{eq:floVsec}]:
\eq{\label{MinkV21}\dot{{V}}_\kappa''= \bigg(1-\frac{\eta_\kappa}{d+2} \bigg) \!\left ( -\frac{  V_\kappa^{(4)}}{Z_\kappa}+ \frac{3 \mbox{${ {V}}_\kappa^{(3)}$}^2}{2 Z_\kappa^2M_\kappa^2} \right )\!\frac{v_d\kappa^{d+2} }{2M_\kappa^3}.}
This is readily seen to derive from the following flow equation for the potential
\eq{\label{eq:UVlimitflotV}\dot V_\kappa&= \bigg(1-\frac{\eta_\kappa}{d+2} \bigg) \frac{v_d\kappa^{d+2} }{M_\kappa} ,}
which generalizes the Minkowski flow obtained in Ref.~\cite{Guilleux:2015pma} in the LPA ($Z_\kappa=1$).
Finally, the prescription \eqn{eq:eta_precsription} yields the following implicit equation for the running anomalous dimension
\eq{\label{eq:Minketa1}
\eta_\kappa &= \bigg(1-\frac{\eta_\kappa}{d+2} \bigg)  \frac{5 v_d\mbox{${ {V}}_\kappa^{(3)}$}^2}{16 Z_\kappa^3}\frac{\kappa^{d+2}}{M_\kappa^7} \Bigg |_{\rm min}.
}
As announced, the expressions \eqn{MinkV21} and \eqn{eq:Minketa1} reproduce the Minkowski results derived in Appendix~\ref{appsec:Mink}.

\subsection{Infrared regime for light fields: Dimensional reduction}

We now consider the regime of interest for our present purposes, namely, the infrared regime $\kappa\ll1$ and the regions of field space where the curvature of the potential $V_\kappa''\ll1$, for which the flow equation \eqn{eq:gamma_dot} can be given a simple analytical form. The calculation is detailed in Appendix~\ref{appsec:IRlimit}.
Our final expression is
\eq{\label{eq:final}
\dot \Gamma_\kappa^{(2)}(\omega) &= \frac{\kappa^{2}}{\Omega_{D+1}M_\kappa^4} \bigg(1-\frac{\eta_\kappa}{2}\bigg)\nn&\times\left (\frac{ V_\kappa^{(4)}}{Z_\kappa} - \frac{2\mbox{${ {V}}_\kappa^{(3)}$}^2}{Z_\kappa^2} \frac{4M_\kappa^2-id\omega}{(2M_\kappa^2-id\omega)^2}  \right ).
}
As before, \Eqn{eq:floVsec} yields the flow of the curvature of the potential 
\eq{
\dot{{V}}_\kappa'' = \bigg(1-\frac{\eta_\kappa}{2}\bigg)\!\left ( - \frac{ V_\kappa^{(4)}}{Z_\kappa} + \frac{2 \mbox{${{V}}_\kappa^{(3)}$}^2 }{ Z_\kappa^2 M_\kappa^2} \right )\!\frac{\kappa^{2}}{\Omega_{D+1}M_\kappa^4} , \label{eq:flowN1}
}
which derives from
\eq{\label{eq:flopothere}
\dot V_\kappa = \bigg(1-\frac{\eta_\kappa}{2}\bigg) \frac{\kappa^{2}}{\Omega_{D+1}M_\kappa^{2}}.
}
The running anomalous dimension is obtained from \Eqn{eq:eta_precsription} as
\eq{
\eta_\kappa =\bigg(1-\frac{\eta_\kappa}{2}\bigg) \frac{3\mbox{${ {V}}_\kappa^{(3)}$}^2}{2Z_\kappa^3}\frac{\kappa^{2}}{\Omega_{D+1}M_\kappa^8}\bigg|_{\rm min} .\label{eq:eta_IR}
}
The flow \eqn{eq:flopothere} of the potential reproduces the result of Ref.~\cite{Serreau:2013eoa} obtained by a direct evaluation of \Eqn{eq:wetterich} with the {\it Ansatz} \eqn{eq:LPAprime} at constant field configuration. It is nothing but the LPA flow corrected by the factor $1-\eta_\kappa/2$ and the explicit dependence on $Z_\kappa$ in the expression of $M_\kappa^2$. As pointed out in this article, \Eqn{eq:flopothere} is similar to the flow equation of the potential in the LPA' in a zero-dimensional Euclidean theory. This results from the dramatic amplification of quantum fluctuations on superhorizon modes, as discussed in detail in the LPA in Ref.~\cite{Guilleux:2015pma}. We shall discuss the expression \eqn{eq:eta_IR} and the role of the anomalous dimension in relation with this effective dimensional reduction in Sec.~\ref{sec:discussion}.

For practical calculations, it is useful to absorb all explicit dependences on the factor $Z_\kappa$ in a field redefinition so that $Z_\kappa$ enters the flow equations only through the running anomalous dimension $\eta_\kappa$. To this aim, we introduce 
\eq{\tilde V_\kappa( \tilde \phi  ) = V_\kappa(\phi)\,,\, \,{\rm with}\quad\tilde\phi=\sqrt{Z_\kappa}\,\phi,}
such that the above flow equations become
\eq{
\label{eq:trululu}\dot{\tilde{V}}_\kappa &= \frac{\eta_\kappa}{2}\tilde V_\kappa'+\bigg(1-\frac{\eta_\kappa}{2}\bigg) \frac{\kappa^{2}}{\Omega_{D+1}M_\kappa^{2}},\\
\label{eq:pouet} \eta_\kappa &=\bigg(1-\frac{\eta_\kappa}{2}\bigg) \frac{3\mbox{${ \tilde{V}}_\kappa^{(3)}$}^2\kappa^2}{2\Omega_{D+1}M_\kappa^8}\bigg|_{\rm min},
}
with $M_\kappa^2=\kappa^2+\tilde V_\kappa''$. We shall discuss the properties and solutions of these equations below. Before doing so, we generalize them to the case of $N$ scalar fields with O($N$) symmetry.

\section{Multiple fields}
\label{sec:multiple}

Generalizing the discussion of the previous sections to an O($N$) theory is straightforward. One must take into account that the coefficients of the derivative expansion are functions of the O($N$) invariant $\rho=\phi_a\phi_a/(2N)$ and that there are various invariant terms at each order in derivatives. For instance, at second order, one has, with an obvious shorthand notation,
\eq{\label{eq:derON}
\Gamma_\kappa [\phi] = -N\!\int_x \left\{U_\kappa(\rho) + \frac{Z_\kappa(\rho)}{2N}(\partial\phi_a)^2 + \frac{Y_\kappa(\rho)}{2}(\partial\rho)^2 \right\}\,\!,
}
where we scaled out various factors $N$ for later convenience. We follow the procedure of Sec.~\ref{sec:derivexp} and compute the transform \eqn{eq:laplace} of the two-point function \eqn{eq:Gamma2} evaluated at constant field. Introducing the decomposition 
\eq{\Gamma^{(2)}_{\kappa,ab} &=  \Gamma^{(2)}_{\kappa,L} P^L_{ab} + \Gamma^{(2)}_{\kappa,T} P^T_{ab},}
with the projectors $P^L_{ab} = \phi_a \phi_b/\phi^2$ and $P^T_{ab}=\delta_{ab}-P^L_{ab}$, we get, leaving the field dependence implicit,
\eq{
\label{eq:trans} \Gamma_{\kappa,T}^{(2)}(\omega) &=-U_\kappa ' +Z_\kappa \alpha_\omega\\
\label{eq:long} \Gamma_{\kappa,L}^{(2)}(\omega) &= -(U_\kappa'+2\rho U_\kappa '') +(Z_\kappa +2\rho Y_\kappa) \alpha_\omega. 
}
The RG flows of the coefficients of the derivative expansion, $U_\kappa$, $Z_\kappa$, $Y_\kappa$, etc., are obtained from the expansion of $\dot\Gamma_{\kappa,T}^{(2)}(\omega)$ in powers of $\alpha_\omega$.

The spirit of the LPA' is to grab the least nontrivial aspect of the derivative expansion while keeping the explicit calculations as simple as possible. In particular, we have seen that a great simplification comes from not having derivative terms in the three- and higher-point functions. This means keeping only derivative terms up to quadratic order in the field, i.e., \Eqn{eq:derON} with $Z_\kappa(\rho)=Z_\kappa$ and $Y_\kappa(\rho)=0$, that is,
\eq{
\Gamma_\kappa^{\rm LPA'} [\phi] = -\int_x \left\{NU_\kappa(\rho) + \frac{Z_\kappa}{2}\partial_\mu\phi_a\partial^\mu \phi_a \right\}. \label{eq:LPAprimeON}
}
Accordingly, we define the running anomalous dimension from the flow of the transverse two-point function\footnote{At the level of the {\it Ansatz} \eqn{eq:LPAprimeON}, it seems equally valid to extract the running anomalous dimension from the flow of the longitudinal component; see \Eqn{eq:long}. This would, however, give different results, which is one of the inconsistencies of the LPA'. The choice \eqn{eq:etatrans} is consistent with the complete derivative expansion at order two. Instead, the flow of the longitudinal two-point function contains information about the function $Y_\kappa$.}
\eq{\label{eq:etatrans}
\eta_\kappa =- \frac{1}{Z_\kappa} \frac{{ \partial} \dot \Gamma_{\kappa,T}^{(2)}(\omega)}{{ \partial}\alpha_\omega} \bigg |_{\omega=0,{ \rm min}}.
}
The steps to computing $\dot \Gamma_{\kappa,T}^{(2)}(\omega)$ are the same as previously. After some straightforward algebra, we first get
\eq{
\dot \Gamma_{\kappa,T}^{(2)}(\eta,\eta',K) &= \frac{1}{2} \tilde \partial_\tau  \int_Q\Bigg\{-   \Big[(N+1)U_\kappa''   G^T_\kappa(\eta,\eta,Q) \label{eq:GflowN}\nn
&+ \left(U_\kappa''+2\rho U_\kappa^{(3)}\right)    G^L_\kappa(\eta,\eta,Q)\Big] \delta_c(\eta,\eta')\nn
&+ 4i\rho{U_\kappa''}^2    G^T_\kappa(\eta,\eta',Q)   G^L_\kappa(\eta,\eta',L) \Bigg \},
}
and
\eq{\label{eq:ONflow}
\dot \Gamma_{\kappa,T}^{(2)}(\omega) = \frac{Z_\kappa\Omega_d}{N(2\pi)^d} & \bigg \{ (N{+}1)U_\kappa'' J_\kappa^T + ( U_\kappa''{+}2\rho U_\kappa^{(3)})  J_\kappa^L \nonumber\\
&+ 2 \rho {U''_\kappa}^2 \sum_{n=0}^4 I_\kappa^{(n)} \bigg \},
}
where $J_\kappa^{T,L}$ and $I_\kappa^{(n)}$ have similar expressions as for \mbox{$N=1$;} see Eqs.~\eqn{eq:Jdef}--\eqn{eq:I_4}. Explicitly,
\eq{
J_\kappa^{T,L} &=  \int_0^{\mathrlap{\kappa}}dp \! \int_p^{\mathrlap{\kappa}} dr\, p^2\!A(p,p,r)\,\hat F^{T,L}_\kappa(p,r) \hat \rho^{T,L}_\kappa(r,p),
}
whereas
\eq{
&I_\kappa^{(0)} =  \int_0^{\mathrlap{\kappa}}dp \int_p^{\mathrlap{\kappa}} dq \int_p^{\mathrlap{\kappa}} dr A(p,q,r)\nonumber\\
&\times\Big\{ \hat \rho^T_\kappa(p,r)\hat F^T_\kappa(r,q) \hat \rho^L_\kappa(q,p)+ \hat \rho^L_\kappa(p,r)\hat F^L_\kappa(r,q) \hat \rho^T_\kappa(q,p)\Big\},}
and similarly for the others: all $I_\kappa^{(n)}$'s can be obtained from Eqs.~\eqn{eq:I_0}--\eqn{eq:I_4} by doubling each line with the structure $TTL + LLT$ in terms of the transverse and longitudinal components of the propagator. 

As before, the flow equation \eqn{eq:ONflow} greatly simplifies in the regimes where the RG scale and the transverse and longitudinal curvatures of the potential are either large or small in units of the space-time curvature. As already mentioned, the first case corresponds to the flat-space limit. Below, we focus on the other limit, where the space-time curvature effects are important.

\subsection{Infrared regime for light fields}

As before, it is useful to absorb explicit dependences on the factor $Z_\kappa$ in a field redefinition. We introduce 
\eq{
U_\kappa(\rho)=\tilde U_\kappa(\tilde \rho)\,,\, \,{\rm with}\quad \tilde \rho = Z_\kappa \rho,
}
as well as the curvatures of the potential in the transverse and longitudinal directions
\eq{
M_{\kappa,T}^2 = \kappa^2 + \tilde U_\kappa', \quad M_{\kappa,L}^2 = \kappa^2 + \tilde U_\kappa'+ 2 \tilde \rho \tilde U_\kappa''.
}
The calculation of the various integrals entering the flow equations in the light infrared limit goes along similar lines as in the case $N=1$. We do not detail it here and refer the reader to Ref. ~\cite{Maximethesis}. Our final expression in the limit  $\kappa^2,M^2_{\kappa,T},M^2_{\kappa,L}\ll1$ reads
\begin{widetext}
\eq{ \dot \Gamma_{\kappa,T}^{(2)}(\omega)  =  \frac{Z_\kappa(2-\eta_\kappa)\kappa^2}{2N\Omega_{D+1}} \Bigg\{ &\frac{(N+1)\tilde{U}_\kappa''}{M_{\kappa,T}^4} + \frac{\tilde{U}_\kappa''+2 \tilde\rho \tilde{U}_\kappa^{(3)}}{M_{\kappa,L}^4}  +  \frac{4 \tilde \rho \tilde{U}_\kappa''^2}{(id\omega-2\tilde \rho \tilde U_\kappa'')M_{\kappa,T}^4} +  \frac{4 \tilde \rho \tilde U_\kappa''^2}{(id\omega+2\tilde \rho \tilde U_\kappa'')M_{\kappa,L}^4}\nn
 & - \frac{ 32id\omega \tilde \rho \tilde U_\kappa''^2}{\big(id\omega + M_{\kappa,T}^2 +M_{\kappa,L}^2\big)^2\left[(id\omega)^2 - \big(M_{\kappa,T}^2-M_{\kappa,L}^2\big)^2\right]}  \Bigg \} . \label{eq:gamma_N} }
\end{widetext}
The flow of $\tilde U_\kappa'$ is obtained by setting $\omega=0$ [see \Eqn{eq:trans}], from which we deduce the flow of the potential 
\eq{\label{eq:flowpotON}
\dot{\tilde{U}}_\kappa=\eta_\kappa\tilde\rho\tilde U_\kappa'+\left(1-\frac{\eta_\kappa}{2}\right) \frac{\kappa^{2}}{N\Omega_{D+1}}\left(\frac{1}{M_{\kappa,L}^{2}}+\frac{N-1}{M_{\kappa,T}^{2}}\right).}
Again, this reproduces the result of Ref.~\cite{Serreau:2013eoa} and assumes the same form as the LPA' flow of a zero-dimensional Euclidean theory. Finally, the running anomalous dimension \eqn{eq:etatrans} obtains as
\begin{equation}
\label{eq:dimanche}
\eta_\kappa = \frac{(2-\eta_\kappa)\kappa^2}{2N \Omega_{D+1}{ \bar \rho_\kappa}}\left[\frac{1}{\kappa^4}+\frac{1}{M_{\kappa,L}^4}-\frac{8}{(\kappa^2+M_{\kappa,L}^2)^2}\right]_{\rm min},
\end{equation}
where we noted $\bar \rho_\kappa$ the minimum of the running potential, defined as $\tilde U_\kappa'(\bar\rho_\kappa)=0$, and we have used $M_{\kappa,T}^2|_{\rm min} = \kappa^2$. We now discuss the properties and the solutions of the LPA' flow equations in the infrared regime, Eqs.~\eqn{eq:trululu}, \eqn{eq:pouet}, \eqn{eq:flowpotON}, and \eqn{eq:dimanche}.

\section{Discussion}\label{sec:discussion}

In this section, we consider the flow equations derived in the previous sections in the LPA' and their solutions in comparison to the results obtained previously in the LPA \cite{Serreau:2013eoa,Guilleux:2015pma}. We focus on the light infrared regime described previously, where the effects of the space-time curvature lead to nonperturbative phenomena.

\subsection{Dimensional reduction}\label{sec:dimred}

As mentioned above, a dramatic consequence of the gravitational amplification of quantum fluctuations on superhorizon scales is that the flow of the potential is effectively turned into that of a zero-dimensional Euclidean theory. This observation originally stems from comparing the flow of the potential in the infrared regime, e.g., \Eqn{eq:flopothere} in the case $N=1$,
\eq{
\dot V_\kappa = \frac{1}{\Omega_{D+1}}\bigg( 1- \frac{\eta_\kappa}{2} \bigg ) \frac{\kappa^2}{\kappa^2+V_\kappa''/Z_\kappa}, \label{eq:IRflowLPAprime}
}
with that of the same theory on the $D$-dimensional Euclidean plane. Using an O($D$) symmetric regulator of the form \eqn{eq:regulator}, the latter reads, in the LPA' \cite{Berges:2000ew,Delamotte:2007pf}, 
\eq{\label{eq:Euclidflow}
\dot V_\kappa = 2v_D \bigg( 1- \frac{\eta_\kappa}{D+2} \bigg ) \frac{\kappa^{D+2}}{\kappa^2+V_\kappa''/Z_\kappa},
}
where $v_D$ has been defined in \Eqn{eq:Minko}. Setting $D=0$ in \Eqn{eq:Euclidflow} yields \Eqn{eq:IRflowLPAprime} up to the numerical prefactor\footnote{Such constant prefactors can always be absorbed in a rescaling of the field and of the potential.} $2v_{D=0}=1\to 1/\Omega_{D+1}$. The same applies to the case $N>1$ as well \cite{Serreau:2013eoa}. 

In the LPA case ($\eta_\kappa=0$), \Eqn{eq:IRflowLPAprime} can also be put in relation with the flow of the potential on the unit $D$-dimensional sphere $S_D$, provided one uses an appropriate $S_D$-symmetric regulator~\cite{Benedetti:2014gja,Guilleux:2015pma}. In that case, the factor $\Omega_{D+1}$ is simply the volume of the compact manifold and the dimensional reduction for $\kappa\ll1$ is a trivial consequence of the discreteness of the spectrum of the corresponding Laplace-Beltrami operator; see also Ref.~\cite{Hu:1986cv}. All nonzero modes, which effectively behave as heavy modes with masses inversely proportional to the sphere radius, decouple and only the (constant) zero mode fluctuations contribute to the RG flow, thereby effectively reducing the dynamics to that of a single degree of freedom.

In principle, the flow equation \eqn{eq:IRflowLPAprime} needs to be supplemented by the expression of $\eta_\kappa$, which is one of the purposes of the present work. But a remarkable consequence of the effective dimensional reduction discussed here is that the LPA' flow \eqn{eq:IRflowLPAprime} can be exactly mapped onto the corresponding one in the LPA ($Z_\kappa=1$). Indeed, we have already mentioned that all explicit  occurrences of the factor $Z_\kappa$ can be traded for extra dependences in $\eta_\kappa$ by a rescaling of the field; see, e.g., Eqs.~\eqn{eq:trululu} and \eqn{eq:pouet}. The dimensionally reduced flow \eqn{eq:IRflowLPAprime} offers an alternative, stemming from the additional symmetry of the $\beta$ function for the potential defined as $\dot V_\kappa=\beta(V_\kappa'',\kappa^2)$:
\eq{
\beta(V_\kappa'',\kappa^2) = \beta(\alpha V_\kappa'',\alpha \kappa^2)\quad\forall \alpha.
}
Choosing $\alpha = Z_\kappa$ and defining $\tilde \kappa= \sqrt{Z_\kappa}\kappa$, it is straightforward to check, using $\kappa \partial_\kappa = (1-\eta_\kappa/2)\tilde \kappa \partial_{\tilde \kappa}$, that
\eq{\label{eq:slowedflow}
\tilde\kappa\partial_{\tilde\kappa}{V}_{ \kappa} = \frac{1}{\Omega_{D+1}} \frac{ \tilde \kappa^{2}}{ \tilde \kappa^{2} + V_{ \kappa}''}.
}
As announced, this is nothing but the LPA flow, which amounts to setting $Z_\kappa=1$ in \Eqn{eq:IRflowLPAprime}. All references to the factor $Z_\kappa$ or its flow $\eta_\kappa$ have been absorbed in a redefinition of the RG scale. The same is obviously true for $N>1$.\footnote{It is worth mentioning that this is peculiar to the LPA', where the deviation from the LPA is all contained in the single factor $Z_\kappa$. It would be interesting, although beyond the scope of this work, to study what happens in the complete second order of the derivative expansion.} 

The immediate consequence is that, assuming we start the infrared flow at a scale $\kappa_0\lesssim 1$, with $Z_{\kappa_0}=1$ so that $\tilde\kappa_0=\kappa_0$, the potential in the LPA' will be the same as that of the LPA at the end of the flow,\footnote{This assumes that $\eta_{\kappa=0}<2$ in order for $\tilde\kappa$ to be well-defined down to $\kappa=0$. This is guaranteed to be the case due to the phenomenon of symmetry restoration (see below), which implies that $\eta_{\kappa=0}=0$.} where $\tilde\kappa=\kappa=0$, i.e.,
\eq{\label{eq:LPA'LPA}V_{\kappa=0}^{\rm LPA'}(\phi)=V_{\kappa=0}^{\rm LPA}(\phi).}
 More generally, as discussed in Ref.~\cite{Guilleux:2015pma}, the solution of the zero-dimensional LPA flow \eqn{eq:slowedflow} can be written in terms of a simple normal integral (as opposed to a functional one). At the end of the flow, $V_{\tilde\kappa=0}$ is simply the Legendre transform of the generating function associated with the following probability distribution for constant field values\footnote{The present discussion assumes field values where the curvature of the potential is small, so that the dimensionally reduced flow is justified.}
\eq{\label{eq:stoch}{\cal P}(\phi)\propto \exp\big\{-\Omega_{D+1}V_{\tilde \kappa_0}(\phi)\big\}.}

As shown in Ref.~\cite{Guilleux:2015pma} in the context of the LPA, this probability distribution coincides with the late time result of the stochastic approach to de Sitter dynamics on superhorizon scales \cite{Starobinsky:1994bd}, provided one uses the appropriate potential in \Eqn{eq:stoch}. In particular, the latter is not the bare potential at the UV scale $\kappa=\Lambda$, but rather the running potential at about the horizon scale, where subhorizon fluctuations have been integrated out. Our present result \eqn{eq:stoch} shows that the LPA' modification to the late time distribution function amounts to a slight change in the actual scale where the appropriate potential is evaluated. But for a given potential at the scale $\tilde\kappa_0$ the final result agrees with the stochastic approach. We conclude that, as far as the effective potential $V_{\kappa=0}(\phi)$ or the corresponding probability distribution ${\cal P}(\phi)$ are concerned, the LPA' gives essentially the same result as the LPA; see also Appendix~\ref{appsec:correlators}. This is a welcome result because the stochastic approach has been shown to correctly capture the leading-order nonperturbative dynamics of superhorizon modes \cite{Tsamis:2005hd,Garbrecht:2013coa}, as confirmed by comparisons with various explicit quantum field theory calculations \cite{Serreau:2011fu,Gautier:2013aoa}. In the context of the NPRG approach, this is a direct consequence of the effective dimensional reduction of the flow of the potential in the infrared regime.

Another consequence of the above discussion is that the main conclusions drawn from the LPA studies of Refs.~\cite{Serreau:2013eoa,Guilleux:2015pma} remain valid. In particular, the phenomena of radiative symmetry restoration and dynamical mass generation along the flow can be described in precisely the same terms as in these LPA calculations. The only effect of the running anomalous dimension is to slow down the flow---or, in other words, to make the RG time tick faster---as compared to the LPA one. This may be of interest if one is to interpret $V_\kappa(\phi)$ as a proxy for the effective potential relevant for dynamics at the scale~$\kappa$. We discuss this effect in the next subsection.

\subsection{Slowing down the RG flow}

\begin{figure}
\includegraphics[width=0.45\textwidth]{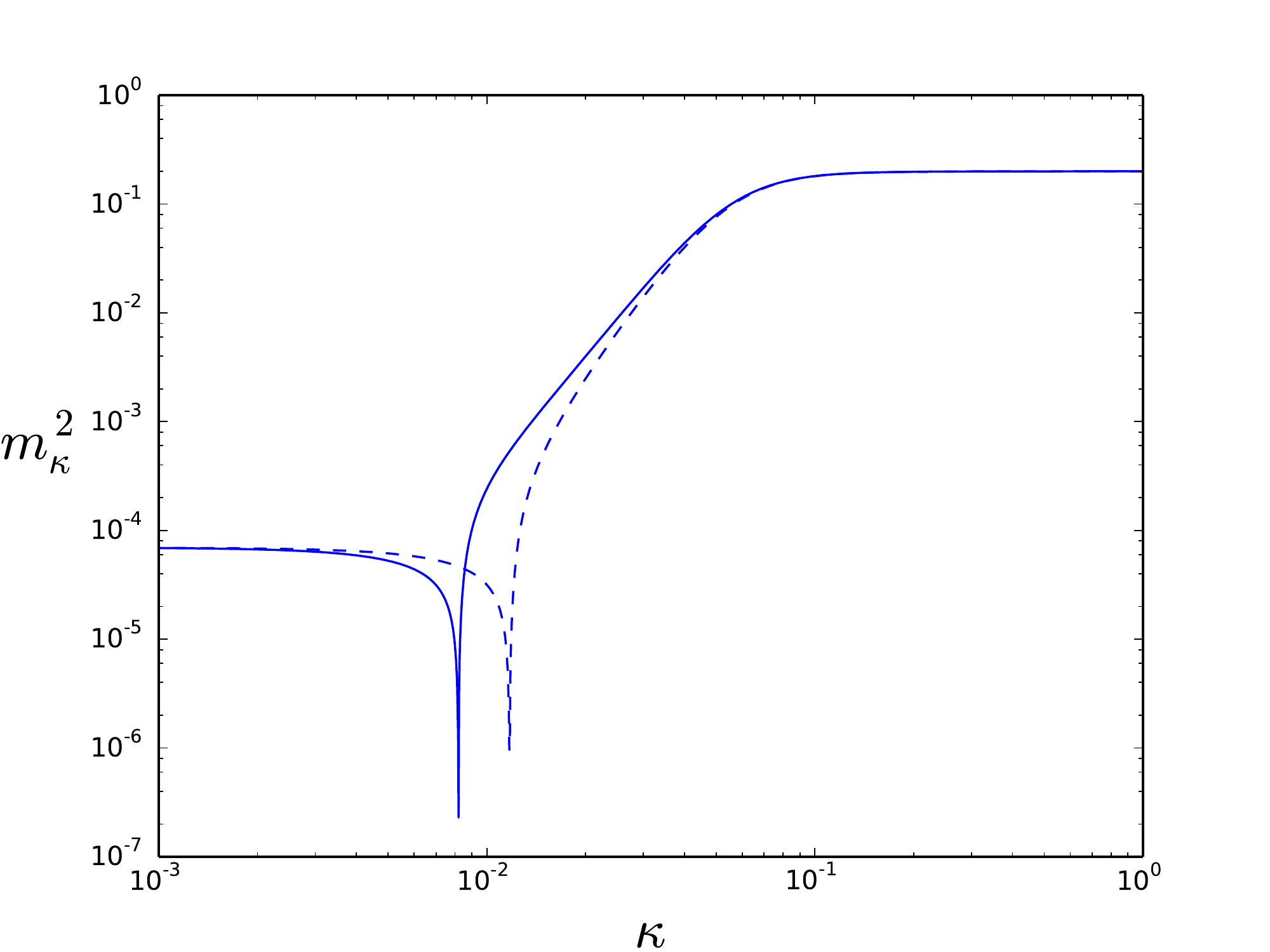}\\
\includegraphics[width=0.45\textwidth]{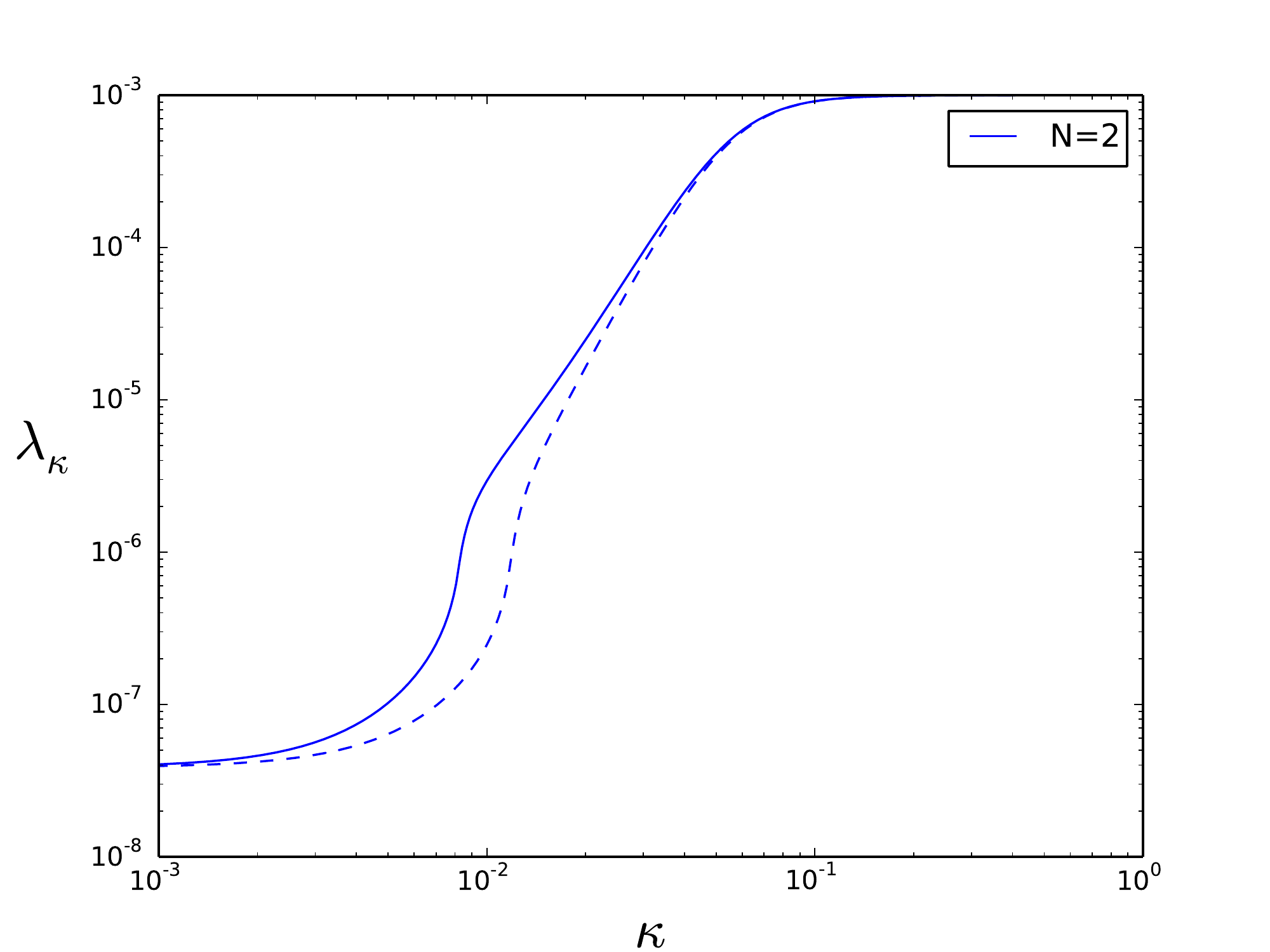}
\caption[delayed LPA' flow]{RG flows of the squared mass (top) and coupling (bottom) in the LPA' (full lines) and LPA (dashed lines) {\it Ans\"{a}tze} for $N=2$ and $D=3+1$. The value at the end of the flow is the same in both cases, though the anomalous dimension slows the flow. Plotting the LPA' flow as a function of $\tilde \kappa = \sqrt{Z_\kappa} \kappa$  coincides exactly with the dashed line.} \label{fig:slowflow}
\end{figure}

We illustrate the actual effect of the running anomalous dimension in the case $N>1$, for which the main features of the flow of the potential can be well captured by a field expansion around the (running) minimum.\footnote{This is not the case for $N=1$ where such a truncation predicts a spurious phase transition \cite{Guilleux:2015pma}.} It is enough to consider the lowest nontrivial order
 \eq{\label{eq:polynomial}
 \tilde U_\kappa(\tilde \rho) = \frac{ \lambda_\kappa}{2}(\tilde \rho-{\bar \rho}_\kappa)^2+{\cal O}\left[(\tilde \rho-{\bar \rho}_\kappa)^3\right].
 }
We initialize the flow at a scale $\kappa_0=1$ with a symmetry breaking potential, i.e., with $\bar\rho_{\kappa_0}>0$. The flows of the parameters $\bar\rho_\kappa$ and $\lambda_\kappa$ are obtained as 
\eq{\dot{ {\bar{\rho}}}_\kappa=-\frac{\dot{\tilde{U}}_\kappa({\bar\rho}_\kappa)}{ \lambda_\kappa}\quad{\rm and}\quad \dot{ \lambda}_\kappa=\dot{\tilde{U}}''_\kappa( {\bar\rho}_\kappa),} respectively. In doing so, one systematically discards higher-order terms, that is, one sets $\tilde U_\kappa^{(n\ge3)}({\bar\rho}_\kappa)=0$. In the light IR regime, see \Eqn{eq:flowpotON}, this yields 
\eq{
&\dot {{\bar \rho}}_\kappa = -\eta_\kappa {\bar \rho}_\kappa + \frac{(2-\eta_\kappa) \kappa^2 }{2 N \Omega_{D+1}} \left[\frac{3}{(\kappa^2+ m_\kappa^2)^2} + \frac{N-1}{\kappa^4} \right ], \label{eq:rhodot} \\
&\dot {\lambda}_{ \kappa} = 2 \eta_\kappa  \lambda_\kappa  + \frac{(2-\eta_\kappa)\kappa^2 \lambda_\kappa^2}{N \Omega_{D+1}} \left [ \frac{9}{(\kappa^2+ m_\kappa^2)^3} + \frac{N-1}{\kappa^6} \right],\label{eq:lambdadot} 
}
where $ m_\kappa^2 = 2  \lambda_\kappa  {\bar \rho}_\kappa$. These equations are valid for $\bar\rho_\kappa>0$ and describe the phenomenon of symmetry restoration: the minimum of the potential $\bar\rho_\kappa$ reaches zero at a finite RG scale and the remaining flow must be described by a expansion around $\tilde\rho=0$; see Ref.~\cite{Guilleux:2015pma}. Figure~\ref{fig:slowflow} shows the numerical integration of these equations up to symmetry restoration, where $m_\kappa^2$ vanishes, followed by the symmetric phase integration, where a nonzero mass develops. The LPA' results are compared to their LPA equivalents. One clearly sees that the LPA' flow is slowed down and that the values at the end of the flow are precisely the same.

\begin{figure}
\begin{center}
\includegraphics[width=0.45\textwidth]{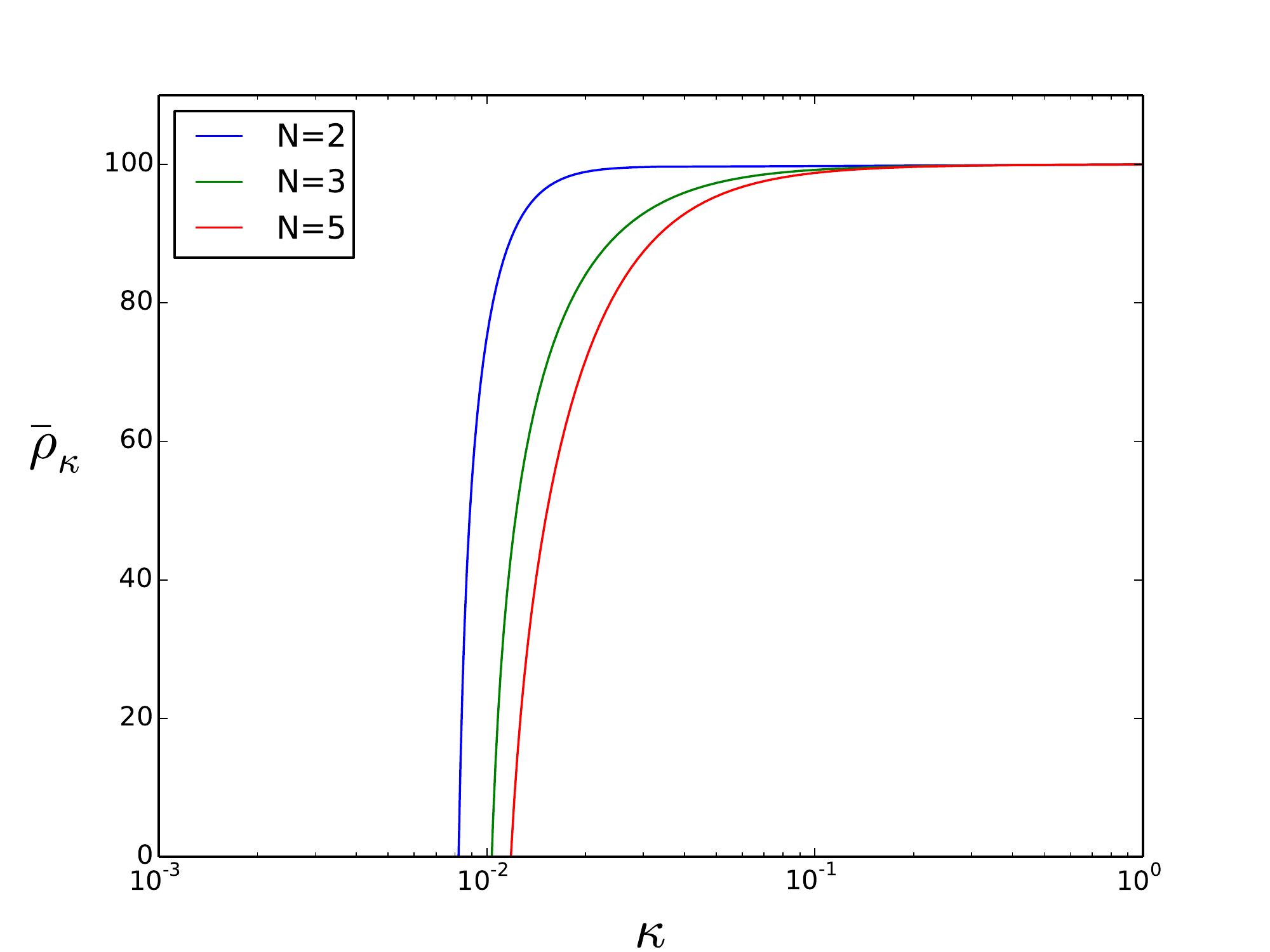}
\includegraphics[width=0.45\textwidth]{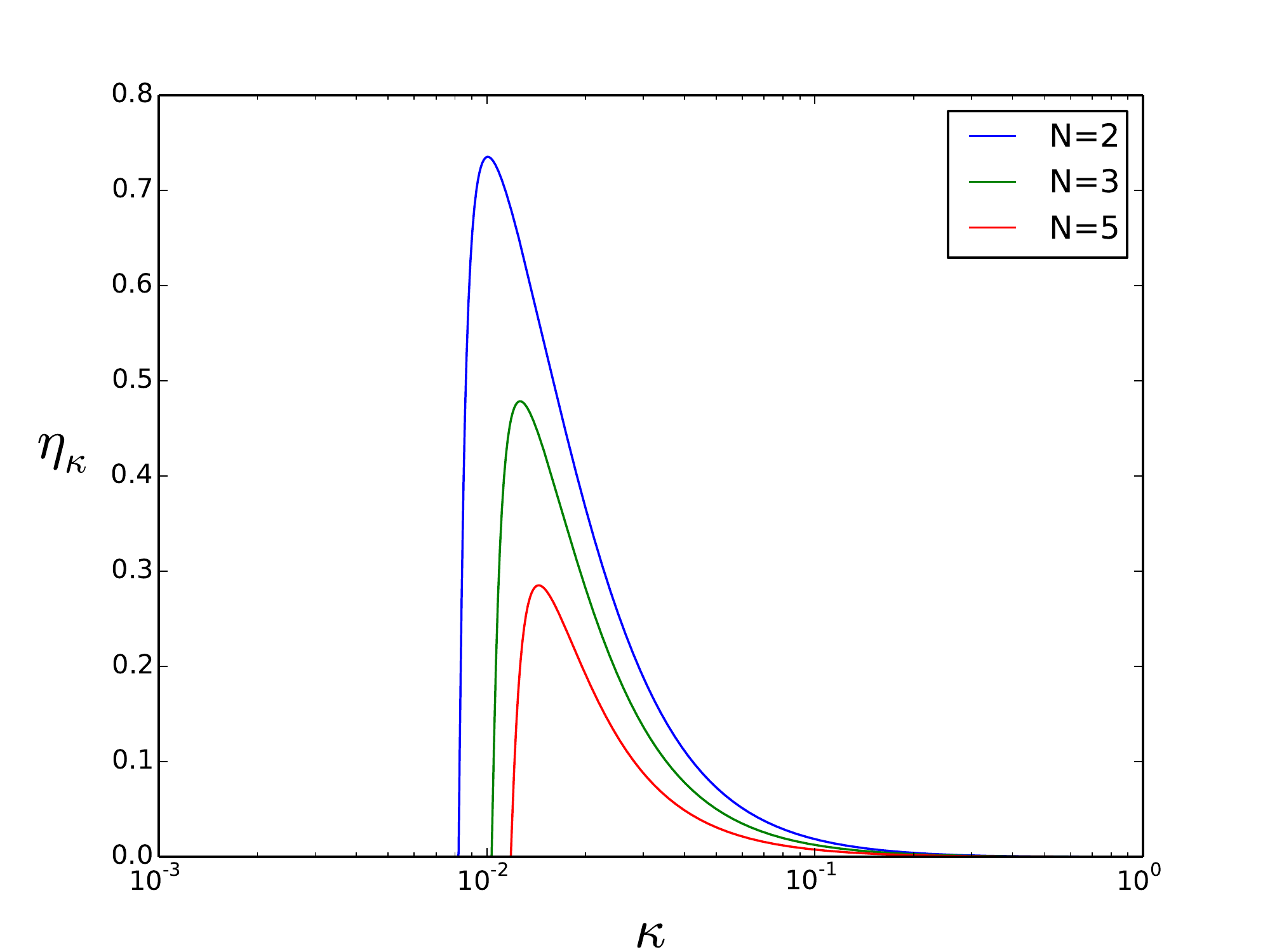}
\caption[anomalous dimension for various $N$]{The top panel shows the infrared flow of the minimum of the potential $\bar \rho_\kappa$ in the LPA' with the polynomial {\it Ansatz} \eqn{eq:polynomial} for various values of $N$. The running anomalous dimension (bottom panel) slows down the flow and delays the symmetry restoration as compared to the LPA result. The effect decreases with increasing $N$ because $\eta_\kappa\propto1/N$. The flow of $\bar\rho_\kappa$ for $N=2$ is qualitatively different from the others and exhibits a clear plateau, where the flow is essentially frozen, followed by an abrupt symmetry restoration.}
\label{fig:flow_rho}
\end{center}
\end{figure}

It is interesting to consider the regime where $m_\kappa^2 \gg \kappa^2$, where the effect of the running anomalous dimension is the most important. In this case, the flow of the potential is completely driven by the Goldstone modes, i.e., the term $\propto N-1$ on the right-hand sides of Eqs.~\eqn{eq:rhodot} and \eqn{eq:lambdadot}, and the running anomalous dimension \eqn{eq:dimanche} approximates to 
\eq{
\eta_\kappa = \frac{2-\eta_\kappa}{2 N \Omega_{D+1}{\bar \rho}_\kappa \kappa^2}. \label{eq:etalargem}
}
Inserting this expression in the flow \eqn{eq:rhodot} of the minimum of the potential yields
\eq{\label{eq:reducedflow}
\dot {{\bar \rho}}_\kappa =  \frac{(2-\eta_\kappa)}{2 N \Omega_{D+1} \kappa^2} \left [ N-2 + O\left (\frac{\kappa^2}{m_\kappa^2}\right) \right ].
} 
This has the same form as the contribution from Goldstone modes in \Eqn{eq:rhodot}, however, with $N{-}1$ replaced by $N{-2}$ in the presence of the running anomalous dimension \eqn{eq:etalargem}. This is due to the $1/\bar \rho_\kappa$ in $\eta_\kappa$, and is reminiscent of the $D=2$ Euclidean XY model.\footnote{There, the same effect together with the appropriate dimensionality generates a line of fixed points. This signals the BKT transition \cite{Berges:2000ew,Delamotte:2007pf}. This does not happen here because the effective dimension is zero.} The flow of $\bar\rho_\kappa$ is thus strongly suppressed for $N=2$, as illustrated in \Fig{fig:flow_rho}. 

In this regime, the flow of the coupling constant is, for $N=2$,
\eq{
\frac{\dot \lambda_\kappa}{\lambda_\kappa^2} \approx \frac{1}{\Omega_{D+1} \kappa^4},
}
which yields, for sufficiently small $\kappa$,
\eq{
\lambda_\kappa \approx 4 \Omega_{D+1} \kappa^4.
}
This regime ends when the square mass $m^2_\kappa\sim\kappa^2$, which happens at a scale
\eq{
\kappa^2 \sim \frac{1}{8 \Omega_{D+1} \bar \rho_\kappa}.
}
where $\bar\rho_\kappa\approx\bar\rho_{\kappa_0}$. At this scale, the running anomalous dimension $\eta_\kappa\sim1$. This regime is illustrated on \Fig{fig:flow_rho}. We observe that symmetry restoration happens abruptly right after exiting this regime of almost frozen flow. 

For $N\ge3$, the effect of the anomalous dimension is less dramatic and decreases with increasing $N$, as $\eta_\kappa\sim1/N$. Explicitly, the LPA' flow of $\bar\rho_\kappa$ gets reduced as compared to the LPA one by the factor $1-\eta_\kappa/2$ and by the change $N-1\to N-2$ in the dominant contribution from Goldstone modes; see \Eqn{eq:reducedflow}.

\section{Conclusion }
\label{sec:conclusion}

In this paper, we have developed the NPRG formalism for quantum fields in de Sitter space beyond the local potential approximation, focusing on the derivative expansion. We have pointed out various difficulties in the very formulation of this expansion as compared to the flat-space case. First, the expansion takes a more complicated form in generally curved space-time as each given order in field derivatives admits {\it a priori} many possible structures associated with the nontrivial curvature. Although this difficulty does not play a role in the maximally symmetric de Sitter space of interest here, it would have to be taken into account for possible applications to, e.g.,  quasi-de Sitter or Friedmann-Robertson-Walker space-times. Second, the formulation of the derivative expansion in terms of a small parameter expansion is a nontrivial task, even in de Sitter space, due to the noncommutation of Killing vectors. As a consequence, we have been unable to express this expansion in terms of the physical momentum, which is the standard procedure in flat space. The solution we have proposed here is, instead, to focus on the time derivatives of spatially homogeneous field configuration. We have then been able to define a systematic prescription for the running parameters of the derivative expansion.

As a first simple, yet nontrivial application, we have implemented the LPA', where one only retains the derivative terms quadratic in the fields. This amounts to a standard kinetic term with a running field normalization factor. We have explicitly computed the flow of the latter---the running anomalous dimension---for O($N$) scalar theories with a simple regulator function and we have studied its effect on the flow of the effective potential as compared to the LPA. It appears that the running anomalous dimension actually takes nonsmall values along the flow, leading to sizable differences with the LPA. However, as in the LPA, the gravitationally enhanced infrared fluctuations effectively reduce the flow of the potential to that of a zero-dimensional theory. As a consequence, the running anomalous dimension merely slows down the flow as compared to the LPA (for identical initial conditions) while the effective potential at $\kappa=0$ is unchanged. In particular, there is no correction to the result of the stochastic approach for the effective potential at the level of the LPA'. 

The slowing down of the flow may still have interesting consequences. Although the potential $V_\kappa$ is, strictly speaking, not physical for $\kappa\neq0$, meaningful quantities can be extracted from the flow. For instance, the scale of symmetry restoration characterizes the maximal size of domains over which a nonvanishing coherent field can develop. It gets modified by the running anomalous dimension. An interesting question concerns the possible implications of the present results both for theoretical studies, e.g., in the context of eternal inflation, and for phenomenological applications, e.g., to inflationary cosmology. 

As already mentioned, the present formalism is based on a closed-time-path (or Keldysh) formulation of the NPRG and uses an infrared regulator function which explicitly breaks Lorentz invariance. This plays no role in the LPA but, in principle, should be taken into account in the LPA'---and, more generally, in the derivative expansion---by introducing different renormalization factors for derivative terms in the temporal and spatial directions, as done in Appendix~\ref{appsec:Mink} in the Minkowski limit. The difficulty is then to compute the running anomalous dimension associated to spatial derivatives. If the anomalous dimensions are not too large, we expect such Lorentz violations to stay under control. This requires further investigation. 

It would, of course, be of interest to go beyond the LPA' and study the complete second order in the derivative expansion. On the technical level, this brings additional complications due to effective derivative interactions vertices in the running action. But this also leads to a richer structure and to potential nontrivial corrections to the LPA/LPA' results and, thus, to the stochastic approach. For instance, the running anomalous dimension is nonzero even in the symmetric regime of the flow, where the minimum of the potential is at $\phi=0$. As a first step in this direction, it would be interesting to actually study the complete second derivative order first in the symmetric regime.  

Another extension of the present work would be to study the formulation of other NPRG approximation schemes in de Sitter space. An interesting example is the Blaizot--Mendez-Galain--Wschebor scheme \cite{Blaizot:2005xy}, tailored to study the full space-time dependence of correlation functions. This could provide a nonperturbative complementary approach to existing studies based on the Dyson-Schwinger equations \cite{Gautier:2013aoa} which are restricted, so far, to coupling or $1/N$ expansions.

Finally, it is of interest to investigate the relation of the present developments with the case of the Euclidean analog of de Sitter space, that is, the $D$-dimensional sphere $S_D$. We have shown in Ref.~\cite{Guilleux:2015pma} that, at the level of the LPA, the effective zero-dimensional flow triggered by the amplified infrared modes in the noncompact de Sitter space is equivalent (using appropriate regulators) to the flow driven by the dynamics of the zero mode on the compact Euclidean sphere. It would be interesting to study the NPRG flow on the sphere $S_D$ beyond the LPA, e.g., using the methods of \cite{Rajaraman:2010xd,Beneke:2012kn,Benedetti:2014gja,Nacir:2016fzi}.

\section*{Acknowledgements}
We are thankful to B.~Delamotte, M.~Tissier, and N.~Wschebor for many useful discussions.

\appendix

\section{Two-point correlators in the LPA'}

\label{appsec:correlators}

Here, we briefly review the derivation of the propagator in the LPA' {\it Ansatz} \cite{Kaya:2013bga,Serreau:2013eoa}. The propagator in the physical momentum representation is defined by the following inversion \cite{Parentani:2012tx}
\eq{\label{appsec:inversion}
\int_{\hat {\cal C}} ds\, \hat G_\kappa^{-1}(p,s) \hat G_\kappa(s,p') = \delta_{\hat {\cal C}}(p-p'),}
with
\eq{
i\hat G_\kappa^{-1}(p,p') = \hat \Gamma_\kappa^{(2)}(p,p') +\frac{\hat R_\kappa(p)}{p^2}\delta_{\hat {\cal C}}(p-p').
}
Here, $\hat \Gamma_\kappa^{(2)}$ is the $p$-represented two-point vertex defined as
\eq{
\Gamma^{(2)}_\kappa(\eta,\eta',K)=K^3\left(\eta\eta'\right)^{\frac{d+3}{2}}\hat \Gamma_\kappa^{(2)}(p,p'),
}
with $p=-K\eta$ and $p'=-K\eta'$. Within the LPA' {\it Ansatz} \eqn{eq:ansatz}, the solution of \Eqn{appsec:inversion} can be written as 
\eq{
\hat F_\kappa(p,p') &= \frac{1}{Z_\kappa} \mathop{\mathrm{Re}} [\hat u_\kappa (p) \hat u_\kappa^* (p')] \\
\hat \rho_\kappa(p,p') &=  - \frac{2}{Z_\kappa} \mathop{\mathrm{Im}} [\hat u_\kappa (p) \hat u_\kappa^* (p')],
}
where the mode function $u_\kappa(p)$ satisfies the equation
\eq{
\left (p^2 \partial_{p}^2  +p^2- \frac{d^2-1}{4}+\frac{V_\kappa''}{Z_\kappa}  \right) \hat u_\kappa (p)  = 0
}
with appropriate boundary conditions. For the regulator \eqn{eq:regulator}, this translates into 
\eq{
p\ge\kappa, &\quad\left( \partial_{p}^2 + 1 - \frac{\nu_\kappa^2 - \frac{1}{4}}{ p^{2}}  \right) \hat u_\kappa (p)  = 0, \\
p\le\kappa, &\quad  \left ( \partial_{p}^2  - \frac{\bar \nu_\kappa^2 - \frac{1}{4}}{ p^{2}}  \right) \hat u_\kappa (p)  = 0, 
}
where $\nu_\kappa=\sqrt{d^2/4-V_\kappa''/Z_\kappa}$ and $\bar \nu_\kappa=\sqrt{\nu_\kappa^2-\kappa^2}$. For boundary conditions corresponding to the Minkowski vacuum in the limit of subhorizon momentum \mbox{$p\to\infty$}, the so-called Chernikov-Tagirov-Bunch-Davies state,\footnote{The normalization is dictated by the equal-time commutation relation for (effective) field operators $[\phi(\eta,{\bf X}),\pi(\eta,{\bf X}')]=i\delta^{(d)}({\bf X}-{\bf X}')/\sqrt{-g}$, where $\pi=\delta{\cal L}_\kappa/\delta(\partial_\eta\phi)$ the canonical conjugate to the field, with the effective Lagrangian scalar density defined from the effective action as $\Gamma_\kappa=\int d^Dx\sqrt{-g}{\cal L}_\kappa$. In the LPA' {\it Ansatz}, we have $\pi=Z_\kappa\partial_\eta\phi$, which leads to the condition on the spectral function $\partial_p\hat\rho_\kappa(p,p')|_{p'=p}=-1/Z_\kappa$. In turn, this gives the Wronskian condition on the mode function $u_\kappa' u_\kappa^*-u_\kappa u_\kappa^{*\prime}=i$} one gets
\eq{
\!\!\!p\ge\kappa, &\,\,\,\, \hat u_\kappa (p) = \sqrt{\frac{\pi p}{4}} e^{i\varphi_\kappa}H_{\nu_\kappa}(p)  \\
\!\!\!p\le\kappa, & \,\,\,\, \hat u_\kappa (p)= \sqrt{\frac{\pi p}{4}} e^{i\varphi_\kappa}\!\left[c_\kappa^+ \left ( \frac{p}{\kappa} \right)^{\!\bar \nu_\kappa} + c_\kappa^- \left ( \frac{\kappa}{p} \right)^{ \!\bar \nu_\kappa}\right] ,
}
where $\varphi_\kappa ={\pi\over2}(\nu_\kappa +{1\over2})$ and $H_\nu(z)$ is the Hankel function of the first kind. The continuity of $u_\kappa (p)$ and $u_\kappa '(p)$ at $p=\kappa$ imposes
\beq
 c_\kappa ^\pm=\frac{1}{2}\left[H_{\nu_\kappa }(\kappa)\pm\frac{\kappa}{\bar\nu_\kappa }H_{\nu_\kappa }'(\kappa)\right].
\eeq
These expressions serve to evaluate the actual form of the flow in the LPA' {\it Ansatz}. 

At this point, it is interesting to notice that the result \eqn{eq:LPA'LPA} can be partly anticipated at the level of the local field variance $\avg{\phi^2(x)}$. Indeed, the above analysis shows that the LPA' propagator is formally that of a free field with mass $V_\kappa''/Z_\kappa$ up to a normalization factor $1/Z_\kappa$: $G_{\rm LPA'}=Z_\kappa^{-1}G_{V_\kappa''/Z_\kappa}$, where $G_{m^2}$ denotes the propagator of a free field of mass $m$. Using the fact that for a light field $\avg{\phi^2}\approx 1/(\Omega_{D+1}m^2)$, we get 
\eq{
\avg{\phi^2}_{\rm LPA'} = \frac{1}{\Omega_{D+1}V_{\rm \kappa=0}''}=\avg{\phi^2}_{\rm LPA}.
}

\section{Infrared limit for light fields}
\label{appsec:IRlimit}

In this section, we give some details concerning the calculation of the integrals $J_\kappa$ and $I_\kappa^{(n)}$, Eqs.~\eqn{eq:Jdef}--\eqn{eq:I_4}, entering the calculation of the LPA' flow. We focus on the limit of infrared scale $\kappa\ll1$ and of small potential curvature $V_\kappa''/Z_\kappa\ll1$. In particular, we show how to simply extract the leading contributions  in inverse powers of the small regulated square mass $M^2_\kappa=\kappa^2+V_\kappa''/Z_\kappa$. A more complete analysis can be found in Ref.~\cite{Maximethesis}. In particular, it is shown there that the integrals $I_\kappa^{(1)}$ and $I_\kappa^{(4)}$, which involve ``mixed'' correlators $\hat G(p,p')$, with $p\le\kappa\le p'$, give subleading contributions. The remaining integrals only involve momenta $p,q,r\le\kappa$ and the leading infrared behavior can be obtained as follows.

First, note that, for $p,p'\le\kappa$, the spectral correlator takes the simple form
\eq{
\hat \rho_\kappa(p,p') = - \dfrac{\sqrt{p p'}}{2Z_\kappa \bar \nu_\kappa} \mysinh{p}{p'}{\bar \nu_\kappa} ,
}
whereas, in the limit $p,p'\le\kappa\ll1$, the  statistical component approximates to
\eq{
\hat F_\kappa(p,p') \approx   \frac{\sqrt{p p'}}{Z_\kappa}\frac{F_{\bar \nu_\kappa}}{(p p ')^{\bar \nu_\kappa}},}
with $F_{\bar \nu_\kappa}=[2^{\bar \nu_\kappa} \Gamma(\bar \nu_\kappa)]^2/(4\pi)= 2^d \Gamma^2(d/2)/(4\pi)+{\cal O}(M_\kappa^2)$. Here, we used the fact that, in the limit we consider here, $\bar\nu_\kappa= d/2-M_\kappa^2/d+{\cal O}(M_\kappa^4)$. Second, as we shall see below, the leading infrared behavior of the relevant integrals come from integrating power laws such as
\eq{
&\int_0^\kappa \frac{dp}{p}\pfrac{p}{\kappa}{\!\varepsilon} = \frac{1}{\varepsilon}, \\
 &\int_0^\kappa \frac{dp}{p} \pfrac{p}{\kappa}{\!\varepsilon}\,\ln \!\left(\frac{\kappa}{p}\right) = \frac{1}{\varepsilon^2}, \label{eq:integrals}
}
where $\varepsilon$ is a small parameter, typically $d-2\bar \nu_\kappa\propto M_\kappa^2$, $i\omega$ or a combination of the two. We shall thus always select terms which give either a small positive power law or a logarithm. With this in mind, we start with the simplest integral, the tadpole contribution.

\begin{widetext}
\subsection{Tadpole contribution}

The tadpole integral reads, explicitly, 
\eq{
J_\kappa &  \equiv \int_0^\kappa\!\!dp \,p^{d-2} \! \int_p^\kappa \frac{dr}{r^2} \hat \rho_\kappa(p,r) \left[ (2-\eta_\kappa)\kappa^2 + \eta_\kappa r^2 \right] \hat F_\kappa(r,p) \\
& = - \frac{F_{\bar\nu_\kappa}}{2\bar \nu_\kappa Z_\kappa^2} \int_0^\kappa \frac{dp}{p}\, p^d \int_p^\kappa \frac{dr}{r} \mysinh{p}{r}{\bar \nu_\kappa}\frac{(2-\eta_\kappa)\kappa^2 + \eta_\kappa r^2}{(pr)^{ \bar \nu_\kappa}}.
}
It is easy to check that the dominant contribution comes from the case where the $r$ integration produces a logarithm. Keeping this one only, we find that, in this limit,
\eq{\label{appeq:J}
J_\kappa & \approx \frac{2^d \Gamma^2(d/2)}{8 \pi \bar \nu_\kappa Z_\kappa^2} (2-\eta_\kappa) \kappa^2 \int_0^\kappa \frac{dp}{p} p^{d-2\bar \nu_\kappa} \ln \left(\frac{\kappa}{p}\right) = \frac{2^d \Gamma^2(d/2)}{8 \pi \bar \nu_\kappa Z_\kappa^2} \frac{2-\eta_\kappa}{(d-2\bar \nu_\kappa)^2} \kappa^{d+2-2\bar \nu_\kappa}.
}

\subsection{Sunset contributions}

We now come to the sunset integrals $I_\kappa^{(n)}$. Just as for the tadpole contribution above, an exhaustive analysis shows that the leading contributions in inverse powers of $M_\kappa$ eventually come from terms where the $r$ integration produces a logarithm. We find
\eq{
I_\kappa^{(0)} &  = \int_0^{\mathrlap{\kappa}}dp \int_p^{\mathrlap{\kappa}} dq \int_p^{\mathrlap{\kappa}} dr \,A_\kappa(p,q,r) \hat \rho_\kappa(p,r)\hat F_\kappa(r,q) \hat \rho_\kappa(q,p) \nn
& \approx \frac{2^d \Gamma^2(d/2)}{16 \pi \bar \nu_\kappa^2 Z_\kappa^3} (2-\eta_\kappa) \kappa^2 \int_0^\kappa \frac{dp}{p} \,p^d \int_p^\kappa \frac{dq}{q} \pfrac{q}{p}{\!\!i\omega} \mysinh{p}{q}{\bar \nu_\kappa}\frac{1}{(pq)^{\bar \nu_\kappa}} \ln \left(\frac{\kappa}{p}\right) \\
I_\kappa^{(2)} &=-  \int_0^{\mathrlap{\kappa}}dp \int_p^{\mathrlap{\kappa}} dq \int_{q}^{\mathrlap{\kappa}} dr\, A_\kappa(p,q,r) \hat F_\kappa(p,r)  \hat \rho_\kappa(r,q) \hat \rho_\kappa(q,p) \nn
 & \approx \frac{2^d \Gamma^2(d/2)}{16 \pi \bar \nu_\kappa^2 Z_\kappa^3} (2-\eta_\kappa) \kappa^2 \int_0^\kappa \frac{dp}{p}\, p^d \int_p^\kappa \frac{dq}{q} \pfrac{q}{p}{\!\!i\omega} \mysinh{p}{q}{\bar \nu_\kappa} \frac{1}{(pq)^{\bar \nu_\kappa}} \ln \left(\frac{\kappa}{q} \right)\\
I_\kappa^{(3)} &=  - \int_0^{\mathrlap{\kappa}}dp \int_p^{\mathrlap{\kappa}} dq \int_p^{\mathrlap{q}} dr\, A_\kappa(p,q,r) \hat \rho_\kappa(p,r)  \hat \rho_\kappa(r,q) \hat F_\kappa(q,p) \nn
& \approx - \frac{2^d \Gamma^2(d/2)}{16 \pi \bar \nu_\kappa^2 Z_\kappa^3} (2-\eta_\kappa) \kappa^2  {\int_0^\kappa\frac{dp}{p}\, p^d \int_p^\kappa \frac{dq}{q}}  \pfrac{q}{p}{\!\!i\omega} \mycosh{p}{q}{\bar \nu_\kappa}  \frac{1}{(pq)^{\bar \nu_\kappa}} \ln \left(\frac{q}{p}\right).
}
On these expressions, one easily checks that only the second terms in the brackets, with a positive power of $q/p$, systematically produce integrals of the type \eqn{eq:integrals}. These terms add up to
\eq{\label{appeq:Is}
I_\kappa^{(0)}+I_\kappa^{(2)}+I_\kappa^{(3)} & \approx -\frac{2^d \Gamma^2(d/2)}{8 \pi \bar \nu_\kappa^2 Z_\kappa^3} (2-\eta_\kappa) \kappa^2 \int_0^\kappa \frac{dp}{p}\, p^d \int_p^\kappa \frac{dq}{q} \pfrac{q}{p}{\!\!i\omega+\bar \nu_\kappa} \frac{1}{(pq)^{\bar \nu_\kappa}} \ln \left(\frac{\kappa}{p}\right) \nn
& = \frac{2^d \Gamma^2(d/2)}{8 \pi \bar \nu_\kappa^2 Z_\kappa^3} \frac{(2-\eta_\kappa) \kappa^2}{i\omega} \int_0^\kappa \frac{dp}{p}\, p^{d-2\bar \nu_\kappa} \left[1-\pfrac{\kappa}{p}{\!\!i\omega}\right] \ln \left(\frac{\kappa}{p} \right)\nn
& = \frac{2^d \Gamma^2(d/2)}{8 \pi \bar \nu_\kappa^2 Z_\kappa^3} \frac{(2-\eta_\kappa) \kappa^{2+d-2\bar \nu_\kappa}}{(d - 2 \bar \nu_\kappa)^2}  \frac{i\omega-2(d - 2 \bar \nu_\kappa)}{(d - 2 \bar \nu_\kappa-i\omega)^2} .
}
Putting together Eqs.~\eqn{eq:gamma_dot}, \eqn{appeq:J}, \eqn{appeq:Is}, and using $d-2\bar\nu_\kappa= 2M_\kappa^2/d+{\cal O}(M_\kappa^4)$, we obtain \Eqn{eq:final}. 
\vspace{.5cm}
\end{widetext}

\section{Flow of the potential}

\label{ap:flow}

In this section, we give the complete expression of the flow of the potential $\dot V_\kappa \equiv \beta_V(V_\kappa'',\kappa)$ in the LPA', which generalizes the expression of Refs.~\cite{Kaya:2013bga,Guilleux:2015pma}. In the case of a single scalar component, it reads
\eq{
\beta(V_\kappa'',\kappa)= A_d \kappa^{d+2}  \big [ (2 - \eta_\kappa) B_d(\nu_\kappa,\kappa)  + \eta_\kappa B_{d+2}(\nu_\kappa,\kappa) \big ],
}
where $A_d = \pi \Omega_d /[8 (2\pi)^d ]$ and 
\eq{
B_d(\nu_\kappa,\kappa) &= \dfrac{e^{- \pi {\rm Im}(\nu_\kappa)}}{d(d^2-4\bar \nu_\kappa^2)} \bigg \{ (d^2 - 2 \bar\nu_\kappa^2) |H_{\nu_\kappa}(\kappa)|^2\nn
& + 2\kappa^2 |H_{\nu_\kappa}'(\kappa)|^2 - 2 d \kappa \,{\rm Re}\!\left[H_{\nu_\kappa}^*(\kappa) H_{\nu_\kappa}'(\kappa) \right] \bigg\} .
}
The generalization to the case $N>1$ is, with the notations of Sec.~\ref{sec:multiple},
\eq{
\dot U_\kappa(\rho) = \frac{1}{N} \left [ (N-1)\beta_V(m_{T,\kappa}^2,\kappa) + \beta_V(m_{L,\kappa}^2,\kappa)\right],
}
where the transverse and longitudinal curvatures are respectively $m_{T,\kappa}^2 = U_\kappa'$ and $m_{L,\kappa}^2 = U_\kappa' + 2 \rho U_\kappa''$.

\section{LPA' in Minkowski space}\label{appsec:Mink}

The results in de Sitter space are expected to have a valid limit $H \to 0$ that should reproduce similar calculations in Minkowski space. To check for this, we perform the analysis of the LPA' flow equation directly in Minkowski space. In particular, we use the formulation of the NPRG on the time contour using the same class of regulators, \Eqn{eq:regreg}. Thus, only spatial momentum directions are regulated, not frequencies. Moreover, because of the symmetries of Minkowski space-time, we can now explicitly study the issue of Lorentz breaking and include independent running anomalous dimensions in the time and space directions. We consider the case $N=1$ for the sake of illustration.

The $D$-dimensional Fourier transform of the (retarded) two-point vertex reads
\begin{align}\label{appsec:MinkLPA'}
\Gamma_\kappa^{(2)}(\omega,K) &\equiv \int_{\cal C} dt'~e^{i\omega(t-t')}\bar \Gamma_\kappa^{(2)}(t-t',K) \nn
&=  -V_\kappa'' + Z_\kappa^{\rm t}\omega^2-Z_\kappa^{\rm s}K^2,
\end{align}
where the second line corresponds to the LPA' {\it Ansatz}.
The flow of this quantity can be directly derived from the Wetterich equation on the contour ${\cal C}$ and is found to be
\eq{\label{appsec:Minkflow}
\dot \Gamma_\kappa^{(2)}(\omega,K) &= \tilde \partial_\tau \int_{q_0,Q} \bigg\{ -\frac{V_\kappa^{(4)}}{2} F_\kappa(q_0,Q) \nn
& \qquad+{V_\kappa^{(3)}}^2 F_\kappa(q_0,Q)G_\kappa^R(\omega-q_0,L) \bigg\},
}
where $L=|{\bf K}-{\bf Q}|$, $\int_{q_0} = \int dq_0/(2\pi)$ and $G_\kappa^R(t-t',Q) =\theta(t-t') \rho_\kappa(t-t',Q) $ is the retarded propagator. Here, we introduced the time Fourier transform
\eq{
f(t)=\int_{q_0}e^{-iq_0t}f(q_0).
}
We use the same notation for the function $f(t)$ and its Fourier transform as there shall be no ambiguity in this section. Using the general arguments of Sec.~\ref{sec:derivexp}, it is easy to check that \Eqn{appsec:Minkflow} is indeed the $H\to0$ limit of the de Sitter flow equation \eqn{eq:flowomega} [see also \Eqn{eq:Gflow}] for the case $K=0$. 

With the LPA' {\it Ansatz} \eqn{appsec:MinkLPA'}, the statistical and retarded correlators can be written as
\begin{align}
F_\kappa(q_0,Q) &= \frac{\pi}{Z_\kappa^{\rm t}} \frac{\delta(q_0 - \bar\omega_Q)+\delta(q_0 +  \bar\omega_Q)}{2  \bar\omega_Q}, \\
 G_\kappa^R(q_0,Q) &= \frac{1}{Z_\kappa^{\rm t}}\frac{1}{\bar\omega_Q^2-(q_0+i\varepsilon)^2} ,
\end{align}
where we define $\bar\omega_Q=\sqrt{z_\kappa}\omega_Q$, with $z_\kappa=Z_\kappa^{\rm s}/Z_\kappa^{\rm t}$, and the regulated frequency
\eq{
\omega_Q = \sqrt{Q^2 + \frac{V_\kappa'' + R_\kappa(Q)}{Z_\kappa^{\rm s}}}.
}
Here, the limit $\varepsilon\to0^+$ is understood.

\subsection{Flow of the potential}

Evaluating $\dot \Gamma^{(2)}(\omega,K)$ at $\omega=K=0$ yields the flow of the effective potential
\eq{
-\dot V_\kappa'' &= \tilde \partial_\tau \int_{q_0,Q} \bigg\{  -\frac{V_\kappa^{(4)}}{2} F_\kappa(q_0,Q) \nn
& +  {V_\kappa^{(3)}}^2 F_\kappa(q_0,Q)G_\kappa^R(q_0,Q) \bigg\}.
}
The frequency integrals are easily computed and we get
\eq{\label{appeq:flotMinkV''}
\dot V_\kappa''= \sqrt{z_\kappa}\,\,\tilde \partial_\tau\! \int_{Q} \bigg(  \frac{V_\kappa^{(4)}}{4Z_\kappa^{\rm s}\omega_Q} -    \frac{{V_\kappa^{(3)}}^2}{8(Z_\kappa^{\rm s})^2\omega_Q^3}\bigg).}
Using the identities
\eq{\tilde\partial_\tau\int_Q\frac{1}{\omega_Q^n}&=-\frac{n}{2Z_\kappa^{\rm s}}\int_Q\frac{\dot R_\kappa(Q)}{\omega_Q^{n+2}}\nn
&=-\frac{n\Omega_d}{d(2\pi)^d}\frac{\kappa^{d+2}}{M_\kappa^{n+2}}\left ( 1- \frac{\eta_\kappa^{\rm s}}{d+2} \right ),
}
where we defined $M^2_\kappa=\kappa^2+V_\kappa''/Z_\kappa^{\rm s}$
and where the second equality uses the regulator 
\eq{\label{appeq:spatialregu}
R_\kappa(Q)=Z_\kappa^{\rm s}(\kappa^2-Q^2)\theta(\kappa^2-Q^2),
} 
we thus obtain
\eq{\dot V_\kappa''=\sqrt{z_\kappa}\frac{v_d\kappa^{d+2}}{2M_\kappa^3}\left ( 1- \frac{\eta_\kappa^{\rm s}}{d+2} \right )\left(-\frac{V_\kappa^{(4)}}{Z_\kappa^{\rm s}}+\frac{3{V_\kappa^{(3)}}^2}{2(Z_\kappa^{\rm s})^2M_\kappa^2}\right),}
which derives from\footnote{Alternatively, this can be obtained more directly by evaluating the flow equation \eqn{eq:wetterich} for constant $\phi$, that is, 
$$
\dot V_\kappa=\frac{1}{2}\int_{q_0,Q}\dot R_\kappa(Q)F_\kappa(q_0,Q)\\
=\sqrt{z_\kappa}\int_Q\frac{\dot R_\kappa(Q)}{4Z_\kappa^{\rm s}\omega_Q},
$$
whose second derivative reproduces \Eqn{appeq:flotMinkV''}.}
\eq{
\dot V_\kappa=\sqrt{z_\kappa}\left ( 1- \frac{\eta_\kappa^{\rm s}}{d+2} \right )\frac{v_d\kappa^{d+2}}{M_\kappa}.
}
As announced, this agrees with the flow \eqn{eq:UVlimitflotV} obtained from the $H\to0$ limit of the de Sitter flow, with $Z_\kappa^{\rm s}=Z_\kappa^{\rm t}=Z_\kappa$.

\subsection{The anomalous dimension in the time direction}

We now compute the running anomalous dimension associated with time derivatives defined as
\eq{
\eta_\kappa^{\rm t}=-\frac{\dot Z_\kappa^{\rm t}}{Z_\kappa^{\rm t}}=\left.-\frac{1}{Z_\kappa^{\rm t}}\partial_{\omega^2} \dot\Gamma^{(2)}_\kappa(\omega,K=0)\right|_{\omega=0}.
}
Evaluating \Eqn{appsec:Minkflow} at $K=0$ yields
\eq{
\dot \Gamma_\kappa^{(2)}(\omega,K=0) &= \tilde \partial_\tau \int_{q_0,Q} \bigg\{  -\frac{V_\kappa^{(4)}}{2} F_\kappa(q_0,Q) \nn
& +  {V_\kappa^{(3)}}^2 F_\kappa(q_0,Q)G_\kappa^R(\omega-q_0,Q) \bigg\}.
}
The frequency integration gives, after setting $\varepsilon\to0$,
\begin{align}
\int _{q_0} F_\kappa(q_0,Q)G_\kappa^R(\omega-q_0,Q) = \frac{1}{2 (Z_\kappa^{\rm t})^2\bar\omega_Q \left(4\bar\omega_Q^2-\omega^2\right)}, 
\end{align}
and we thus get, after derivation with respect to $\omega^2$ at $\omega=0$,
\begin{align}
\eta_\kappa^{\rm t}=-\frac{\sqrt{z_\kappa}\,{V_\kappa^{(3)}}^2}{32(Z_\kappa^{\rm s})^3} \tilde \partial_\tau \int_Q \frac{1}{\omega_Q^5} = \frac{5\sqrt{z_\kappa}\,{V_\kappa^{(3)}}^2}{64(Z_\kappa^{\rm s})^4}  \int_Q \frac{\dot R_\kappa}{\omega_Q^7}.
\end{align}
This expression is valid for any regulator function $R_\kappa$. With the choice \eqn{appeq:spatialregu}, we obtain
\eq{\label{appeq:timeandim}
\eta_\kappa^{\rm t} = \sqrt{z_\kappa}\left ( 1- \frac{\eta_\kappa^{\rm s}}{d+2} \right )\frac{5v_d\mbox{${\tilde  {V}}_\kappa^{(3)}$}^2\kappa^{d+2}}{16(Z_\kappa^{\rm s})^3M_\kappa^7},
}
which, again, agrees with the $H\to0$ limit of our de Sitter calculation (for $z_\kappa=1$); \Eqn{eq:Minketa1}. This is what we expect, since our definition of the anomalous dimension in the de Sitter case indeed corresponds to the time direction. 

\subsection{The anomalous dimension in the spatial direction}

In the Minkowski case, there is no difficulty to unambiguously extract the renormalization factor of the spatial gradients through an expansion in $K^2$. The corresponding running anomalous dimension is defined as 
\eq{
\eta_\kappa^{\rm s}=-\frac{\dot Z_\kappa^{\rm s}}{Z_\kappa^{\rm s}}=\left.\frac{1}{Z_\kappa^{\rm s}}\partial_{K^2} \dot\Gamma^{(2)}_\kappa(\omega=0,K)\right|_{K=0}.
}
Evaluating \Eqn{appsec:Minkflow} at $\omega=0$ yields
\eq{
\dot \Gamma_\kappa^{(2)}(\omega=0,K) &= \tilde \partial_\tau \int_{q_0,Q} \bigg\{-\frac{  V_\kappa^{(4)}}{2} F_\kappa(q_0,Q)\nn
& +  {V_\kappa^{(3)}}^2 F_\kappa(q_0,Q)G_\kappa^R(q_0,L) \bigg\},
} 
where the frequency integral is evaluated as
\begin{align}
 \int_{q_0} F_\kappa(q_0,Q)G_\kappa^R(q_0,L) = \frac{ \Omega_{Q,L}^2}{ 2(Z_\kappa^{\rm t})^2\bar\omega_Q\left(\Omega_{Q,L}^4+4\varepsilon^2 \bar\omega_Q^2\right)},
\end{align}
where $\Omega_{Q,L}^2 = \bar \omega_L^2-\bar\omega_Q^2+\varepsilon^2$. Here, the limit $\varepsilon\to0$ must be taken with care because evaluating the $K^2$ derivative at $K=0$  involves terms $\sim1/\varepsilon^2$. 

Using the following identity for a given function $g(Q,L=|{\bf K}-{\bf Q}|)$,
\begin{align}
&\partial_{K^2} \int_Q g(Q,L) \bigg |_{K=0} \nn
&= \int_Q \left[\partial_{L^2}g(Q,L) + \frac{2Q^2}{d}\partial_{L^2}^2g(Q,L) \right]_{L=Q},
\end{align}
we obtain, after rescaling $\varepsilon\to\sqrt{z_\kappa}\varepsilon$,
\begin{align}\label{appeq:dotZspatial}
\dot Z_\kappa^{\rm s} =  \frac{\sqrt{z_\kappa}\,{V_\kappa^{(3)}}^2}{2(Z_\kappa^{\rm s})^2\varepsilon^2} \tilde \partial_\tau&\! \int_Q \frac{1}{\omega_Q}  \Bigg\{\frac{4}{d}Q^2(1+r_\kappa')^2 \frac{ 12\omega_Q^2-\varepsilon^2}{(4\omega_Q^2+\varepsilon^2)^3} \nonumber \\
&- \left(1+r_\kappa'+\frac{2Q^2}{d}r_\kappa''\right) \frac{4\omega_Q^2-\varepsilon^2}{(4\omega_Q^2+\varepsilon^2)^2} \Bigg\},
\end{align}
where we introduced the function $r_\kappa(Q^2) = R_\kappa(Q)/Z_\kappa^{\rm s}$. 

To proceed, it is useful to note the identity
\begin{align}\label{appeq:identity}
& \tilde \partial_\tau\int_Q \frac{1}{\omega_Q^n} \left \{ \left(1+r_\kappa'+\frac{2Q^2}{d}r_\kappa''\right)-\frac{nQ^2}{d\omega_Q^2}(1+r_\kappa')^2  \right\} \nn
&= \tilde \partial_\tau\int _Q\frac{1}{\omega_Q^n} \left \{ 1-\frac{nQ^2}{d\omega_Q^2}(z_\kappa+r_\kappa')  \right\} \nn
&= \tilde \partial_\tau\left [ \frac{\Omega_dQ^d}{d(2\pi)^d\omega_Q^n} \right ]_0^\infty  =0.
\end{align}
The second line follows from integrating by part the contribution involving $r_\kappa''$, whereas the last equality uses the fact that the regulator vanishes in the limit of infinite momentum. Using this identity for $n=3$, one shows that the terms $\sim 1/\varepsilon^2$ in \Eqn{appeq:dotZspatial} vanish identically, as expected. We thus have, for $\varepsilon \to 0$,
\eq{
\dot Z_\kappa^{\rm s} = \frac{\sqrt{z_\kappa}\,{V_\kappa^{(3)}}^2}{16(Z_\kappa^{\rm s})^2} \tilde \partial_\tau &\int_Q \frac{1}{\omega_Q^5} \Bigg\{   -\frac{5Q^2}{d\omega_Q^2}(1+r_\kappa')^2 \nn
& \quad+\frac{3}{2} \left(1+r_\kappa'+\frac{2Q^2}{d}r_\kappa''\right) \Bigg\} .
}
Using the identity \eqn{appeq:identity} with $n=5$, this rewrites as
\eq{
\dot Z_\kappa^{\rm s}=  \frac{{5}\sqrt{z_\kappa}\,{V_\kappa^{(3)}}^2}{{ 32}d(Z_\kappa^{\rm s})^2 } \tilde \partial_\tau \int_Q  \frac{Q^2 }{\omega_Q^7}(1+r_\kappa')^2.
}
Finally, expressing the operator $ \tilde \partial_\tau$ explicitly, we obtain
\begin{align}
\dot Z_\kappa^{\rm s}= \frac{{ 5}\sqrt{z_\kappa}\,{V_\kappa^{(3)}}^2}{{ 16}d(Z_\kappa^{\rm s})^3 } \!\int_Q  \frac{Q^2}{\omega_Q^7} \left \{\dot R_\kappa' (1+r_\kappa') - \frac{{ 7}\dot R_\kappa(1+r_\kappa')^2}{4\omega_Q^2} \right \}\!,
\end{align}
valid for any function $R_\kappa$. With the choice \eqn{appeq:spatialregu}, we have
\eq{
\dot R_\kappa(Q)&=Z_\kappa^{\rm s}\left[(2-\eta_\kappa^{\rm s} )\kappa^2+\eta_\kappa^{\rm s} Q^2\right]\theta(\kappa^2-Q^2),\\
\dot R_\kappa'(Q)&=Z_\kappa^{\rm s}\eta_\kappa^{\rm s} \theta(\kappa^2-Q^2)-2Z_\kappa^{\rm s}\kappa^2\delta(\kappa^2-Q^2),
}
and $1+r_\kappa'(Q^2)=\theta(Q^2-\kappa^2)$, with $\theta(0)=1/2$. This yields
\begin{align}
\eta_\kappa^{\rm s} = \sqrt{z_\kappa} \,\frac{5 v_d\mbox{${ {V}}_\kappa^{(3)}$}^2\kappa^{d+2}}{16(Z_\kappa^{\rm s})^3M_\kappa^7} .
\end{align}
As expected, this differs from the anomalous dimension \eqn{appeq:timeandim} in the time direction. This is a consequence of the breaking of the Lorentz symmetry through our regulator function. The difference
\eq{
\eta_\kappa^{\rm s}-\eta_\kappa^{\rm t}=\frac{\left(\eta_\kappa^{\rm s}\right)^2}{d+2}
}
is a measure of the violation of Lorentz symmetry along the flow. Clearly, the latter is small when the anomalous dimensions themselves are small. For instance, at the Wilson-Fisher fixed point in $D=4-\epsilon$, where the anomalous dimensions are ${\cal O}(\epsilon^2)$ we have $\eta^{\rm s}_{\rm WF}=\eta^{\rm t}_{\rm WF}=5\epsilon^2/54+{\cal O}(\epsilon^3)$. For comparison,  the same LPA' exercise in Euclidean space with a fully O($D$) invariant regulator of the form \eqn{eq:regulator} yields $\eta_{\rm WF}=\epsilon^2/12+{\cal O}(\epsilon^3)$.

\end{document}